\documentclass{llncs}

\usepackage{graphicx}
\usepackage{stmaryrd}
\usepackage{geometry}
\usepackage{dsfont}
\usepackage{amssymb}
\usepackage[english]{babel}
\usepackage[utf8]{inputenc}
\usepackage{tikz}
\usepackage{tipa}
\usepackage{setspace}
\usepackage{multicol}

\makeatletter
\def\captionof#1#2{{\def\@captype{#1}#2}}
\makeatother

\begin{document}

\title{ On the Complexity of Computing Minimal Unsatisfiable LTL Formulas }

\author{Francois Hantry\inst{1}  \and Lakhdar Sa\"is\inst{2} \and Mohand-Saïd Hacid\inst{1} }
\date{}
\institute{Universit\'e Claude Bernard Lyon 1, LIRIS CNRS UMR 5205, France
{\email{$\lbrace$mohand-said.hacid,francois.hantry$\rbrace$@liris.cnrs.fr}}
\and 
Universit\'e d'Artois,\\ CRIL CNRS UMR 8188, France\\
{\email{sais@cril.univ-artois.fr}
}
}

\maketitle

\begin{abstract}
  We show that (1) the Minimal False QCNF search-problem (MF-search) and the Minimal Unsatisfiable LTL formula search problem (MU-search) are FPSPACE complete because of the very expressive power of QBF/LTL, (2)  we extend the  PSPACE-hardness  of the MF decision problem to the MU decision problem. As a consequence, we deduce  a positive  answer to the open question of PSPACE hardness of the inherent Vacuity Checking problem. We even show that the  Inherent Non Vacuous formula search problem is also FPSPACE-complete.

\end{abstract}


\section{Introduction}

Recently, the notion of Minimal Unsatisfiable Linear Temporal Logic formula (MU for LTL) has been introduced in \cite{Schuppan10}. This notion is, for instance,  fundamental  to reduce the search space in LTL sat-solvers \cite{CimattiRST07}, \cite{Hantry11F}, or to understand the cause of unsatisfiability and enable debugging \cite{Schuppan10}, \cite{Raman2011}, \cite{Hantry11F}. Intuitively, an element  $g \in MU (f)$ from a LTL unsatisfiable formula $f$ is a limit weakening\footnote{some substitutions by $TRUE$ (resp.$FALSE$) of some subformula occurrences of positive (resp.negative) polarity} of $f$ such that $g$ remains unsatisfiable. 
We consider the following two fundamental problems:
\begin{multicols}{2}
\begin{center}
{\bf  MU-decision problem}\\
{\bf input}: a LTL formula $f$\\
{\bf output}: yes while $f$ is minimal unsatisfiable, no otherwise.\\

{\bf  MU-search problem}\\
{\bf input}: a LTL formula $f$\\
{\bf output}: $g \in MU(f)$ while $f$ is  unsatisfiable, no otherwise.
\end{center}
\end{multicols}
 The aim of this work is to study computational complexity of the above MU-decision/search problems. The authors of \cite{Papadimitriou1988} have shown  that the $MU_{CNF}$ decision problem is $D^P$-complete for propositional logic with formula in Conjunctive Normal Form (CNF)  but it is in P while   the deficiency is fixed \cite{FleischnerKS02}.  An important effort has been devoted to approaches allowing to approximate/compute $MU_{CNF}$ of propositional logic (see,\cite{ZM03},\cite{LynceM04},\cite{LiffitonS09},\cite{Nadel10},\cite{SilvaL11}).  The Minimal False QBF decision problem is PSPACE complete \cite{BuningZ06} but it is in $D^P$  for fixed deficiency \cite{BuningZ08}. However, only a few investigations dealt with MF \cite{YuM05}.  The author in \cite{Schuppan10} defines MU for LTL, and recalls that given a formula and a fixed occurrence, deciding if it is not necessary (w.r.t unsatisfiability) is PSPACE complete (MU-step-dec). However, the MU decision/search problems (MU-dec/search) remain open.  A few work propose computation of MU for LTL \cite{CimattiRST07},\cite{Schuppan10},\cite{Raman2011}, \cite{Hantry11F}.  A recent work also proposed to compute  minimal revision of an unsatisfiable LTL specification \cite{Fainekos11} in order to achieve the satisfiability. Some simple results for unrealizability of LTL formula (2-EXPTIME complete) are  also given in \cite{Schuppan10}. In \cite{BeerBCOT09}, the authors investigated the causes in counterexample of LTL specification and have shown that the decision problem is NP complete (by considering as inputs a LTL counter-example, a timestamped variable and a LTL formula). This work is built on the theory of causes introduced in \cite{HalpernP01}. Also  based on the theory  of \cite{HalpernP01}, the authors in \cite{ChocklerHK08} analyze  some variables as a cause of verifying a model checking test.       In \cite{BelovS11}, the authors investigated basic algorithms for Minimal Unsatisfiable boolean circuit. Computing the minimal unsatisfiable formulas  in SMT is proposed in \cite{CimattiGS07}. Since a Minimal Unsatisfiable LTL formula is a particular case of inherent non  vacuity \cite{FismanKSV08},\cite{Schuppan10}, we consider also complexity result for inherent vacuity. Given a LTL Formula $f$ and a  fixed subformula occurrence $Occ(g)$, deciding  if $Occ(g) $ is  a witness of  inherent vacuity of $f$ is PSPACE-complete   but deciding whether there is an inherent vacuity in $f$  is still an open problem \cite{FismanKSV08}.  While $f\in LTL $ is a conjunction, the decision problem of a smallest equivalent subset of the $f$'s conjuncts (irredundancy)   and of a given size  is PSPACE complete \cite{ChocklerS09}.  Some works were devoted to the vacuity detection (see, \cite{BeerBER01},\cite{KupfermanV03},\cite{ArmoniFFGPTV03}, \cite{Namjoshi04},\cite{SimmondsDGC07},\cite{FismanKSV08},\cite{ChocklerS09}). To summarize, although substantial complexity results have been provided in the propositional case, current corresponding complexity results for LTL  appear to be less studied than in the propositional case. Mainly,   complexity results for minimality problems in the LTL case assume additional subformula or  length parameter in the definition of the problem. In this paper, we show that (1) the Minimal False QBF search-problem (MF-search) and the Minimal Unsatisfiable LTL formula search problem (MU-search) are FPSPACE complete because of the very expressive power of QBF/LTL and (2)  we extend the  PSPACE-hardness  of the MF decision problem to the MU decision problem. As a consequence, we provide  a positive  answer to the open question of PSPACE hardness of the inherent Vacuity Checking problem. We even show that the   Inherent Non Vacuous formula search problem is also FPSPACE-complete. 

 For uniformity purpose we introduce QLTL ($QBF \subset QLTL$ and $LTL \subset QLTL$) in   Section 2. We also discuss the notions of weakening for QLTL formulas and minimal unsatisfiable LTL formulas.  In Section 3,  we start by  analyzing the complexity of Minimal FALSE QCNF formulas, then, we enhance the translation of QCNF sat  to LTL Model Checking to show complexity results  for the LTL MU-search problem. Finally, we  propose an original proof of PSPACE completeness of the MU-dec problem.  We reuse these results in order to provide complexity results for Inherent Vacuity Checking. We conclude in Section 4. 








\section{Preliminaries}
{\bf Complexity}\\
We recall the basic definition of computational complexity \cite{papadimitriou},\cite{StockmeyerM73}. Let $\Sigma$ be an alphabet, a total  deterministic computable function $f$ from $\Sigma^* $ to $\Sigma ^* $,  associated with a total (at left) binary relation $R(x,y) \in \Sigma ^* \times \Sigma ^*$   with input $x\in \Sigma^*$  is an always accepting deterministic Turing Machine with  three tapes: a `two-ways' `read-only' input tape (where $x$ lies), a   `two-ways' `read/write' computing tape and a   `one-way'`write-only' output tape with output $f(x)$ such that $R(x,f(x))$.
A FPSPACE search  problem with relation $R$   is such that there exists a polynomial $P$ such that $R(x,y) \Rightarrow |y| \leq P(|x|)$ and there exists a function $f$ such that for any $x$,  the total use of space units of the machine (of the output and of the working tapes) is also  bounded by   $P(|x|)$.  A  decision problem associated to a language $L \subset \Sigma ^ *$ over a fixed alphabet $\Sigma$ is PSPACE iff there exists a FPSPACE function $f$ such that  given any input $x \in \Sigma^* $, $f(x)= `yes'$ iff $x \in L$ (thus for instance $f(x)=`no'$ while $x \notin L$). A logspace function is a function $f$  using $\mathcal O(log(|x|)$ space at  the working tape but no constraint at the ouptut tape\footnote{However, one can show that the output is polynomial in $|x|$ yet.}. There exists a logspace reduction  of a decision problem  $L_1 \subseteq \Sigma^ * $ to another $L_2\subseteq \Sigma ^*$  iff   there exists a  particular  logspace function  $f$ such that $x \in L_1 $ iff $f(x) \in L_2$. There exists a logspace reduction from  a relation $R_1$  to a relation $R_2$ iff there exists three functions $f$, $g_2$, $h$ with  function $f$ and $h$ are logspace functions, and $g_2$ is a $R_2$ function such that for any $x$, $R_1(x,h(g_2(f(x)),x))$ holds, i.e., given a $x$ one can compute a $y$ with $R_1(x,y)$ by (1) computing $f(x)$, (2) computing $z=g_2(f(x))$ with $R_2(f(x),z)$ and (3) computing $y=h(g_2(f(x)),x)$.  A PSPACE decision problem is PSPACE-complete iff any PSPACE-problem is logspace reducible to it. A FPSPACE search problem is FPSPACE complete iff any FPSPACE problem is logspace reducible to it. We quickly recall that a NP decision problem is a decision problem which is solvable by a Non-deterministic Turing Machine, in polytime for the positive answer. $D^P$ is the class of languages of the form $L_1 \cap L_2$ with   $L1$ a NP problem and  $L_2$ a Co-NP problem (Intuitively one positive call and one negative call to a NP-complete problem). $\Sigma_2 ^P$ is the set of decision problems with a non deterministic polytime Turing Machine, but with a NP-complete oracle. We recall that $NP \subseteq D^P \subseteq \Sigma_2 ^P \subseteq PSPACE$, without knowing whether the inclusions are strict.  The reduction from one of the non deterministic problem is usually through a deterministic polytime computable function rather than a logspace function.

{\bf QLTL \cite{SistlaVW87}}

 Let $P$ be a non empty finite set of propositional variables, $p \in P$ and $A$ and $B$ are two QLTL formulas. A temporal logic formula is inductively built by means of the following rules:
\begin{center}
TRUE \textbar FALSE \textbar
 $ p$ \textbar $A \wedge B $ \textbar $A \vee B $ \textbar $\neg A$\textbar $\mathcal X (A)$\textbar
  $A \, \mathcal U B$ \textbar $A\, \mathcal W B$ \textbar $\exists p \, A$ \textbar $\forall p \, A$ .
\end{center} Furthermore,  $\mathcal G (A) = A\, \mathcal W  \, \mathrm{FALSE}$ and $\mathcal F (A) = \mathrm{TRUE} \, \mathcal{U} \, A$. In this paper, while some definitions hold for QLTL formulas, the focus is on two fragments of QLTL: Quantified Boolean Formula (QBF \cite{StockmeyerM73}) and Linear Temporal Logic (LTL \cite{Emerson}). QBF is the fragment of QLTL without modal operators ($\mathcal U, \mathcal W, \mathcal X, \mathcal F, \mathcal G$) and LTL is the fragment without quantifiers ($\exists , \forall$).  Both the satisfiability and Model Checking decision problems of LTL and QBF are PSPACE complete on the contrary to the satisfiability problem of QLTL which is non-elementary\footnote{Elementary is the class of decision problems for which the execution time is bounded by a finite composition of exponential in the input size $|x|$ (e.g., $\mathcal O (2^{2^{2^{|x|}}})$ unit times).} and the model checking is however PSPACE complete \cite{PanV06}.  The set of QBF without quantifier is denoted PROP and is NP-complete \cite{Cook71}. A QLTL formula is in Prenex form iff  it is of the form $ Q  \overline x \phi$ with $Q\overline x = Q_1 x_1 Q_2 x_2...Q_n x_n$ with $Q_i \in \{\forall ;\exists\}$, and $\overline x=( x_1,...,x_n)$ standing for a set of different variables, and $\phi$ without  quantifier. In the following, we will assume that any QLTL is in Prenex Form. Except for the case of Vacuity checking (see Section 3), we will also restrict  any formula to  possibly  contain $\neg$ symbol  solely applied to propositional variable(s) \cite{Fisher91}. We call such a formula Negative Normal Form (NNF).\\
 A propositional variable $p$ in a QLTL $f$ is free iff there exists an occurrence of $p$ in $f$ which is not in the scope of a quantifier. A closed  QLTL is a QLTL without free variables.  A literal $ l$  is either a propositional variable $p \in P$, or its negation $ \neg p$. $lit(P)$ denotes the set of literals of $P$. We\footnote{This definition is necessary because $\neg \neg p $ is syntactically different from $p$.} define $\sim$ (1) on literals as $\sim p = \neg p $ and $\sim \neg p =p$; (2) on quantifiers as  $\sim \forall = \exists$ and $\sim \exists = \forall$ and $\sim (Q_1   Q) x_1 \overline x   = \sim (Q_1) x_1 \sim( Q)\overline x$. A clause is a disjunction of literal(s). A QLTL-clause is a disjunction of literal(s) and/or modal operator(s) applied to literal(s) (e.g., $(a \, \mathcal U \neg b ) \vee \neg c \vee \mathcal F (d)$). We finally say that $\Phi \in QBF$ is a QCNF if it is of the special prenex form $Q \overline x \phi$ with $\phi$ a conjunction of different clause(s). In this case, if $\Phi$ gets no quantifier then it lies in $CNF\subset PROP$. Note that the QCNF-sat decision problem is also PSPACE complete by adapting the proof of \cite{StockmeyerM73} to the QCNF case. By analyzing the proof of the PSPACE-hardness of QBF in \cite{StockmeyerM73},  one can also show  that the QBF-sat-search problem is FPSPACE-complete. This is the problem of searching  a satisfiable valuation of the free variables of the QBF formula while it is satisfiable. To prove the FPSPACE hardness, let consider the following points. Since  the output tape is PSPACE bounded in the definition of a FPSPACE problem, the configurations can also contain output tape variables. This is then sufficient to remove the external existential quantifiers of the configurations in  the proof of \cite{StockmeyerM73} to prove QBF-sat search is FPSPACE-hard. The inclusion in FPSPACE is trivial.   \\
A  linear time structure is an element $\mathcal{M}$ in $(2 ^P) ^\mathbb{N}$. $\forall i  \in \mathbb{N}$,$\forall \mathcal{M}$$ \in (2 ^P) ^\mathbb{N}$:
\begin{itemize}
\item $ (\mathcal{M},i) \vDash p$ with $p \in P$ iff $p \in \mathcal{M}(i)$.
\item  $ (\mathcal{M},i) \vDash \mathcal X (A)$ iff  $ (\mathcal{M},i+1) \vDash A$.
\item  $ (\mathcal{M},i) \vDash A\, \mathcal UB$ iff  $ \exists j \geq i, (\mathcal{M},j) \vDash B$ and $\forall k ,  i \leq k <j, (\mathcal{M},k) \vDash A $.
\item  $ (\mathcal{\mathcal{M}},i) \vDash A\mathcal WB$ iff  $\forall j \geq i, (\mathcal{M},j)  \vDash A$ or ( $\exists j \geq i, (\mathcal{M},j) \vDash B$ and $\forall k ,i \leq k <j, (\mathcal{M},k)\vDash A  $).
\item The semantics of any propositional combination is defined as usual.
\item  $ (\mathcal{M},i) \vDash \mathcal \exists p \, (A)$ iff  there exists a linear structure $\mathcal M  '$ such that  $(\mathcal{M}',i) \vDash A$ and  where $\mathcal M'$ differs  from  $\mathcal M$  solely at the instances of $p$. 
\item  $ (\mathcal{M},i) \vDash \mathcal \forall p \, (A)$ iff  for any  linear structure $\mathcal M  '$ such that  $\mathcal M'$ differs  from  $\mathcal M$  solely at the instances of $p$ then  $(\mathcal{M}',i) \vDash A$. 
\end{itemize}

A partial instance is a linear structure where solely some variables are instantiated (at any state).\\ 
We write down $\mathcal {M} _t$ for the suffix of $\mathcal M$ starting at time $t$.\\

A Kripke Structure $\mathcal K$ is a labeled automaton $\mathcal K = (S,S_0,T,l)$ with $S$ the set of states, $S_0\subset S$ the set of initial states, $T \subset S\times S$ a total binary relation standing for the transitions and $l$ a total function from $S$ to $2^P$. A $\mathcal K$-linear structure   is any linear structure $\mathcal M$ such that there exists a function $m$ in $S^\mathbb N$ such that $m(0)\in S_0$ and $\forall i\geq 0$  $\mathcal M (i) =l(m(i))$ and $(m(i),m(i+1)) \in T$.  We note $\mathcal K \vDash f$ iff any $\mathcal K$-linear structure $\mathcal M$ is such that $ (\mathcal{M},0) \vDash f$. In this paper we restrict ourselves to finite Kripke Structure. It may happen that a state $s$ occurs in a formula, without confusion, it stands for the conjunction of its literals $\wedge_{p \in l(s)} p  \wedge_{p \in P\setminus l(s)}\neg  p$.\\
Let $f$ be a QLTL formula, a syntactic tree $T(f)$ is defined by the following rules:

\begin{itemize}
\item $T(p \in P)$ is a single  node labeled by  $p$. 
\item $T(\circ g)$ is a tree with a  root node which is labeled by $\circ$ ($\circ \in \{ \neg ; \mathcal X ; \exists ; \forall \}$) and a child subtree $T(g)$. 
\item $T(g_1 \circ g_2)$ is a tree with a root node which is labeled by $\circ$ ($\circ \in \{\mathcal U ; \mathcal W ; \vee; \wedge \}$) with a left child subtree $T(g_1)$ and a right child subtree $T(g_2 )$.
\end{itemize}
A subformula $h$ of $f$ is a `subword' of the `word' $f$ such that  $h$ is a also a formula, and  the set of subformulas is denoted $sf(f)$.  We  will also  write $Cl(sf(f))$  the set of clauses which are in $sf(f)$.

\fbox{
 \begin{minipage}[c][5cm] {3.5cm}
\begin{center}

\begin{tikzpicture}
\node {$ \mathcal U  $}
child {node {$\wedge$}
child{node{$a$}}
child {node{ $b$}}
      }
child {node {$\neg$}
	child {node {$a$}}
      }
;
\end{tikzpicture}
\end{center}
\captionof{figure}{\caption{Syntactic tree} }
\label{tree}
\end{minipage}}
\hfill
 \begin{minipage}[c]{10cm}
 For instance $sf((a\wedge b) \mathcal U  \neg a ) = \{ a ; b; a \wedge b ; \neg a ;  (a \wedge b )\mathcal U  \neg a   \}$. The set of subformula occurrences $Occ(sf(f))$ corresponds to the set of nodes of $T(f)$. For each node $N$, a natural subformula $Sf(f)(N) $ can be associated with the subformula of the $N$-root subtree of $T(f)$. For instance  on Figure 1, $Occ(sf((a\wedge b) \mathcal U  \neg a))$ gets two occurrences of the subformula $a$. 
Furthermore $a \wedge b$ is associated with the labeled node $\wedge$. We also define $Cl(Occ(sf(f)))$ as before.  Let $g$ be  another QLTL  formula,  $f[h\leftarrow g]$ is the result of the substitutions in $f$  of all the occurrences of $h$ by $g$.
For one specific occurrence of $h$ denoted $Occ(h) \in Occ(sf(f))$,  $f[Occ(h)\leftarrow g]$ is the result of the substitution in $f$ of the only occurrence $Occ(h)$  of $h$ by $g$. 
\end{minipage}

We divide $Occ(sf(f))$ into two disjoint sets: $Occ(sf(f))=sf^+(f) \cup sf^-(f)$ where $sf^+(f)$ is the set of the subformula occurrences with positive polarity\footnote{A subformula occurrence with positive polarity is a subformula occurrence which is in the scope of an  even number of negation(s). The negative case corresponds to an odd number of negation(s).}. We fix $lit^\epsilon(f)=  sf^\epsilon(f)  \cap lit(P)$ with $\epsilon \in \{+;-\}$. Finally if $Occ(g) \in Occ(sf(f))$ then $Occ(g') \in Occ(sf(f))$ is a superformula occurrence of $Occ(g)$ iff  $Occ(g)$ is a descendant node of $Occ(g')$ in $T(f)$.  For instance on Figure 1, $\neg a $ is a  superformula occurrence of the `second' occurrence  $a$ in $ \neg a $. If $\mathcal K$ is a Kripke structure and $s$ a state of $\mathcal K$, then for any $\mathcal K$-linear structure $(\mathcal M,m)$ different from $S^ * s^ \omega$, $\mathcal M [s\leftarrow erase]$ is the modified $(\mathcal M,m)$ where any corresponding  occurrence of $s$ has been erased.  \\
We call weak promise $wp$ any occurrence of subformula of an QLTL formula $f$ of the form $ (A\, \mathcal UB)$ or $ (A\mathcal WB)$ (we recall $\mathcal F (B)= (TRUE)\, \mathcal UB$),   with $B \neq FALSE$ which is called a promise operand. We  will say that a timestamped state $(i,m(i))$ of a $\mathcal K$ linear structure $\mathcal M$    triggers a weak promise $wp$ iff  $f= Z \wedge \mathcal G (C \Rightarrow \mathcal X ^k[((wp \vee H) \circ D) \wedge E)])$ with $\circ \in \{\mathcal U;\vee\}$, $k \in \mathbb N$  and such that    $(\mathcal M ,i) \vDash C  \wedge \mathcal X ^k(\neg D \wedge  \neg H )$.
 We will say that a timestamped state $(i,m(i))$ of a $\mathcal K$ linear structure $\mathcal M$ propagates or postpones a weak promise $wp$ iff (1) there exists $i'$ with $0 \leq i' \leq i$ such that $(i',m(i'))$ triggers $wp$  and (2) either $(\mathcal M ,j) \vDash A \wedge \neg B$ where $k\leq i-i'$, for any $i'+k\leq j \leq i$, with $B$ the promise operand  of $wp$, or $k > i - i'$. We finally say that a weak promise $wp$ is fulfilled at $(i,m(i))$ iff  there is a $i'$ with $i'\leq i$ where $wp$ is triggered and propagated until $i$ where  $(\mathcal M ,i) \vDash  B$ with  $B$ the promise operand of $wp$.\\





{\bf Weakening  QLTL formulas and Minimal Unsatisfiable QLTL formulas}

For (Quantified) propositional logic, a basic weakening is essentially defined as the deletion of a clause in QCNF \cite{BuningZ06}. It is extended  in \cite{BuningZ07a} as the substitution of a particular `maximal' $\vee$-subformula (a disjunction) occurrences by $TRUE$ while the formula is in QBF $\cap$ NNF. However, for Linear Temporal Logic and related Model Checking, a basic  weakening is usually defined for any subformula occurrence \cite{Schuppan10}. In what follows, 
 we compare these various definitions and describe which occurrences are necessary and sufficient to consider in order to check the minimality of an unsatifiable formula. 

\begin{definition} ( Basic clausal weakening for QCNF)\\
Let  $f =Q\overline x( C_1 \wedge ...\wedge C_m)$ an element of $QCNF$ with the clauses $C_{j}$. Then a basic clausal  weakening of  $f$ is $f[C_{j_0}\leftarrow TRUE]$ for some $j_0 \in [1;m]$.
\end{definition}

For $f_1$, $f_2$  and $f$ in $QCNF$ the relation of basic clausal weakening $R_{Cl(sf)}$ is such that $R_{Cl(sf)}(f_1,f_2)$ iff $f_1$ is a basic weakening of $f_2$. If $R^* _{Cl(sf)}$ is the reflexive, transitive  closure of $R_{Cl(sf)}$, then  the set of weakened subformulas of $f$  is $W_{Cl(sf)}(f) = \{g \in QCNF | R^* _{Cl(sf)}(g,f) \}  $.
\begin{definition}(Basic Occurrence Weakening  in QLTL \cite{Schuppan10})
Let $f \in QLTL $,  a  basic occurrence weakening is a formula $g$ such that  $g$ is the result of a substitution in $f$ of either (1) a subformula  occurrence in $sf^+(f)$  by $TRUE$, or (2) a subformula  occurrence in $sf^-(f)$  by $FALSE$  .
\end{definition}
 For instance if $f=\mathcal G (\neg c \vee( a \vee ((\neg b)\, \mathcal U	c)))$ then $g= \mathcal G (\neg c \vee TRUE)$ is a basic occurrence weakening of $f$.  Except for case of  vacuity checking  (see section 3), we will restrict  ourselves to occurrences in $sf^+$.  $R_{sf^+}$ and $W_{sf^+}$ are defined similarly as  $R_{Cl(sf)}$ and $W_{Cl(sf)}$. 
However, while the occurrence  of $f_1= x \vee (c \wedge ((\neg b) \mathcal W e)) $ is substituted by $TRUE$ in $f=\exists x (\neg r \wedge (r \vee (\mathcal F (x \vee (c \wedge ((\neg b) \mathcal W e)) )))$ then the resulting formula $f'=  \exists x (\neg r \wedge (r \vee (\mathcal F (TRUE )))$ is trivially equivalent to $f''=\exists x (\neg r \wedge ( TRUE))$.  Consider $Eq_0 \in (Sf^+(f))^2$, with  $Eq_0(Occ(f_1),Occ(f_2))$ iff $Occ(f_2)$ is  a superformula occurrence of $Occ(f_1)$  and $f_2$   gets one of the following forms  $f_1 \vee Z $,$A \, \mathcal U /\mathcal  W (f_1)$, $(f_1)\mathcal W FALSE$,  or $\mathcal X (f_1)$. Then, if $Eq$ is the symmetric, reflexive, transitive closure of $Eq_0$, and if $Eq(Occ(f_1),Occ(f_2))$, then $f[Occ(f_1) \leftarrow TRUE] \equiv  f[Occ(f_2) \leftarrow TRUE]$.  A class representative of a class $Cla$ from $Eq$ can be the (right)\footnote{For subformula occurrence there is only one right maximal element}  maximal element of $Cla$ with respect to  $Eq_0^*$. For the last example $Cla(Occ(f_1))=Cla(Occ(f_2))$  with $Cla(Occ(f_2))=(r \vee (\mathcal F (x \vee (c \wedge ((\neg b) \mathcal W e)) ))$ and $Cla(Occ(f_2))$ is maximal. It is then sufficient to consider solely  a `maximal' representative per class for weakening analysis as for the QBF case \cite{BuningZ07a}. But if the maximal class representative or a minimal class representative  is a conjunction or of the form $A\mathcal U / \mathcal W B$ with $A \neq TRUE$ and $B \neq FALSE$, then it is not correct to say its substitution by $TRUE$ is a basic weakening, since any of its conjunct/$A$ substitution by $TRUE$ weakens $f$ `less' than the conjunction or $A\mathcal U / \mathcal W B$.    For instance, if $f= \exists b [( c \wedge \neg d) \vee b \vee d]$, then solely the occurrences  $\{c; \neg d ; \}$ are the `weakest' maximal non-conjunctives occurrences. Similarly, if $f=\exists x (\neg r \wedge (r \vee (\mathcal F (x \vee (c \wedge ((\neg b) \mathcal W e)) )))$ then $\{\neg r \, ; \,  c \, ;\, \neg b \}$ are the weakest maximal non-conjunctive occurrences.      We then define a weakest basic weakening of maximal non-conjunctive subformula occurrences ( Weakest-Max weakening, for short) as follows.

\begin{definition} (Weakest-Max weakening for QLTL)\\
A weakest  basic weakening of maximal non-conjunctive subformula occurrences, is a  basic  Weakening of `maximal ' non-conjunctive subformula occurrence $Occ$, where $Occ$ is the maximal representative element  w.r.t. $Eq^*_0$ of $Cla(Occ)$, and $Cla(Occ)$ does not contain any  Maximal/minimal  element which is a conjunction or of the form $A\mathcal U / \mathcal W B$ with $A \neq TRUE$ and $B\neq FALSE$.
\end{definition}   

$WeakestMAX(sf ^+)(f)$, $R_{WeakestMAX(sf ^+)}$ and related weakened formulas $W_{WeakestMAX(sf ^+)}(f)$ are defined as previously.


\begin{definition}(Minimal Unsatisfiable QLTL Formula)
Let $O$ be the   mapping from any formula $f \in QLTL$ to a set $O(f) \subseteq sf^+(f)$. A  QLTL formula $g$ is Minimal Unsatisfiable w.r.t. $O$ ($ g \in MU_{O}$) iff (1) $g$ is unsatisfiable, (2) $g$  gets no  unsatisfiable proper weakened subformula w.r.t $R_{O}$ (i.e. $W_O(g)\cap UNSAT =\{g\}$). If $f$ is an unsatisfiable QLTL formula, then $MU_O(f)=MU_O \cap W_O(f)$.
\end{definition}
For instance, if $f= \alpha \wedge \neg \alpha \wedge  \mathcal F (o)\wedge \mathcal G (\neg c)\wedge \mathcal G (o \Rightarrow (\mathcal F (p)\wedge \mathcal F (g)))\wedge (\neg g)\mathcal W  p \wedge \mathcal F (i)\wedge (\neg i)\mathcal W  p \wedge \mathcal G (p \Rightarrow \mathcal G (\neg i))$ then  $MU_{sf^+}(f) = \{ \alpha \wedge \neg \alpha \wedge TRUE ; TRUE  \wedge \mathcal F (i) \wedge (\neg i)\mathcal W  p \wedge \mathcal G (p \Rightarrow \mathcal G (\neg i))\}$. Also note that the set $MU_O(f)$ is identical (by simplifying any  $TRUE \vee...$, $\wedge_i TRUE$, $A \mathcal U/ \mathcal W TRUE$ or $\mathcal G (TRUE)$  by $TRUE$) whatever the   $O$ be from our two precedent definitions of weakening ($O \in \{sf^+;WeakestMax(sf^+)\}$). Thus, in the following, we will solely write MU instead of $MU_{O}$. If $f$ is closed, then the unsatisfiability becomes Falsity and we call minimal FALSE (MF) instead of MU. In the remaining part of this paper, the MU-dec/search problem is restricted to the elements of LTL, and the MF-dec/search problem is restricted to the elements of QCNF\footnote{In this case $WeakestMax(sf^+)$ and $Cl(sf)$ are identical }.

\section{Complexity results}
The MU-dec  problem is obviously  in $PSPACE^{PSPACE} =PSPACE$.  To show the hardness  one adapts the proof of hardness for MF-dec \cite{BuningZ06} to LTL.
As a corollary this shows the PSPACE hardness of the Inherent Vacuity decision problem. To show the FPSPACE-hardness of MU-search, we start by showing the FPSPACE-hardness of MF-search in QCNF, then we enhance a QCNF sat / LTL Model Checking reduction from \cite{DemriS98}. We conclude that the inherent non vacuous search problem (INV-search) is FPSPACE-complete.


{\bf Minimal False Formula in QCNF} \\
  We need two lemmas to prepare the proof. W.l.g. , we fix $O=Cl(sf)$.
The first one has been proved in \cite{BuningZ06} but it is recalled to understand its extension later.

\begin{lemma} \label{split} \cite{BuningZ06} Assume $\Phi = \forall y Q \overline x \phi$ is in MF. Then  either  only $y \in lit^+(\Phi)$ occurs  or   only $\neg y \in lit^+(\Phi)$ occurs in $\phi$.\end{lemma}

{\bf (proof)} $\Phi = \forall y Q \overline x\phi  $  is FALSE iff $\Phi[y\leftarrow TRUE] \wedge \Phi[y\leftarrow FALSE]$ is FALSE iff $\Phi[y\leftarrow TRUE]$ is FALSE or $\Phi[y\leftarrow FALSE]$ is FALSE. For instance if $\Phi[y\leftarrow TRUE]$ is FALSE, then  if a clause $C$ containing $y \in lit^+ (\Phi)$ is in $\phi$ this clause can be substituted by $TRUE$ and $(\Phi[y\leftarrow TRUE])[C[y\leftarrow TRUE]\leftarrow TRUE]=(\Phi[C\leftarrow TRUE])[y\leftarrow TRUE]$ remains FALSE. Then $\Phi[C\leftarrow TRUE]$ is FALSE. However, it contradicts the assumption $\Phi$ is in MF. We conclude that there is no  occurrence of $y \in lit^+(\Phi)$  while $\Phi[y\leftarrow TRUE]$ is FALSE. The other case is similar.

\begin{lemma}\label{QCNF}
Let $\Phi =Q \overline x \phi$ be a QBF  in Prenex Form. $\Phi $ is LOGSPACE reducible to an equivalent QCNF denoted $QCNF(\Phi)$.
\end{lemma}

{\bf (Proof)}  Let $Set=\{x_ {\phi } \}$ be the starting set with $x_ {\phi }$ a fresh variable, and $UCS=\emptyset$ the starting  set of  clauses. $\Phi$ is LOGSPACE reducible to an equivalent QCNF by applying the following rules until reaching a fixpoint:

\begin{itemize}
\item If $x_\psi = x_{\psi_1 \wedge \psi_2} \in Set$ then   $\forall j \in \{1;2\}$ $UCS:=\{x_\psi \Rightarrow x_{\psi_j} \} \cup UCS$ and $\forall j $ $Set := Set \cup \{x_{\psi_j} \}.$ 

\item If $x_\psi = x_{\psi_1 \vee \psi_2} \in Set $ then    $UCS:=\{x_\psi \Rightarrow (x_{\psi_1} \vee x_{\psi_2}) \} \cup UCS$ and $\forall j \in \{1;2\} $ $Set := Set \cup \{x_{\psi_j} \}.$ 

\item If  $x_\psi \in Set$ is  such that $\psi \in lit(P) \cup \{TRUE;FALSE\}$ then $UCS:=\{x_\psi \Rightarrow \psi \} \cup UCS.$



\end{itemize}

Let $\overline{x'}$ be a vector standing for the set $ Set$ and  $\phi' = x_{\phi} \wedge_{C \in UCS} C$, then $QCNF(\Phi)=Q \overline x \exists \overline {x'} \phi'\equiv \Phi$. 


\begin{theorem}\label{MFFPSPACE} \label{MF}( MF-Search) Given  $\Phi$ a closed  QCNF, providing a MF of $\Phi$ if $\Phi$ is FALSE, and answer `no' if $\Phi$ is $TRUE$ is a FPSPACE complete problem.
\end{theorem}
{\bf (proof)}
The inclusion in FPSPACE is rather obvious. Let $\Phi_0 = Q\overline y \phi_0$ a QBF in prenex form with free variables $\overline x =(x_1,....,x_n)$. Then $\Phi_0$ is satisfiable iff $\forall \overline x \neg \Phi_0=\forall \overline x (\sim Q) \overline y \neg \phi_0 $ is FALSE. According to lemma \ref{split}, if $\Psi= \forall \overline x (\sim Q) \overline y \exists \overline z \psi$ is a MF of $QCNF(\forall \overline x \neg \Phi_0)=  \forall \overline x (\sim Q) \overline y \exists \overline z  \phi' $, then there exists a corresponding partial instance $\mathcal I_{(\Psi,\overline x)}$  of the $\overline x$
deduced from  $\Psi$ such that $\Psi[\overline x \leftarrow  \mathcal I_{(\Psi,\overline x)}]$ is FALSE. 
 Moreover $\phi' \vDash \psi $ and then we deduce  $  (\sim Q) \overline y \neg \phi_0 [\overline x \leftarrow  \mathcal I_{(\Psi,\overline x)}] \equiv(\sim Q)\overline y \exists \overline z \phi'[\overline x \leftarrow  \mathcal I_{(\Psi,\overline x)}] \vDash (\sim Q) \overline y \exists \overline z  \psi[\overline x \leftarrow \mathcal I_{(\Psi,\overline x)}   ] \equiv FALSE$. This means that $TRUE \equiv \neg[ (\sim Q) \overline y \exists \overline z  \psi[\overline x \leftarrow  \mathcal I_{(\Psi,\overline x)}   ]] \vDash  Q \overline y  \phi_0[\overline x \leftarrow  \sim \mathcal I_{(\Psi,\overline x)}]  $, i.e.,   $\sim \mathcal I_{(\Psi,\overline x)}\vDash \Phi_0$. Thus, finding a satisfiable model of a QBF is $LOGSPACE$ reducible to the search problem of a MF of a closed QCNF. This proves the FPSPACE hardness \cite{papadimitriou}.\\
{\bf Deciding Minimal Unsatisfiable LTL formula }\\
 W.l.g. we solely consider $O =WeakestMax(sf^+)$.
 \begin{lemma}(Definitional SNF\cite{Fisher91}\label{DSNF}) Any LTL formula $f$ can be LOGSPACE reduced to an equi-satisfiable formula in (F,X)-TL \cite{SistlaC85} of the form  $f'= x_{f} \wedge_{2 \leq i \leq m} f' _i$ where any $f' _i$ is one of the following "globally" scoped LTL-clauses-based forms : $\mathcal G (x \vee \mathcal F (x')) $, $ \mathcal G (y \vee \mathcal X (y'))$ or $ \mathcal G (w \vee w' \vee (\delta \wedge w'')) $. The $x_f,x,x',y,y',w,w',w''$  are literals and $\delta \in \{TRUE ;FALSE\}$\footnote{Once $\delta$ is instantiated, the formula is simplified to the equivalent clause}. Furthermore,  no pair of literals in the scope of a $\mathcal G$ operator  have the same propositional variables. Finally, $\wedge_{2 \leq i \leq m} f' _i$ is satisfiable with a model $\mathcal M' $ which sets $x_f$ to FALSE at $\mathcal M'(0)$. 
\end{lemma}

{ \bf (Proof)}\\
 Let the starting set $Set=\{x_f  \}$ with $x_f$ a fresh variable, and $UCS=\emptyset$ the starting  set of unwound LTL-clauses. Let us apply the following rules until reaching a fixpoint:
\begin{itemize}
\item If $x_\psi = x_{\psi_1 \wedge \psi_2} \in Set$ then   $\forall j \in \{1;2 \}$ $UCS:=\{x_\psi \Rightarrow x_{\psi_j} \} \cup UCS$ and $\forall j \in \{1;2 \} $ $Set := Set \cup \{x_{\psi_j} \} $ 

\item If $x_\psi = x_{\psi_1 \vee \psi_2} \in Set $ then    $UCS:=\{x_\psi \Rightarrow (x_{\psi_1} \vee x_{\psi_2}) \} \cup UCS$ and $\forall j \in \{1;2 \} $ $Set := Set \cup \{x_{\psi_j} \} $ 



\item If $x_\psi = x_{\mathcal X (\psi_1)} \in Set $ then 
$UCS:=\{x_\psi \Rightarrow \mathcal X (x_{\psi_1}) \} \cup UCS$ and    $Set := Set \cup \{x_{\psi_1} \} $ 
\item  If  $x_\psi = x_{\psi_1 \mathcal U/ \mathcal W \psi _2} \in Set$ then  $UCS:= \{x_\psi \Rightarrow  (x_{\psi_2} \vee (x_{\psi_1 \wedge \mathcal X (\psi)})) \} \cup  UCS$ and $Set := Set \cup  \{ x_{\psi_2} ; x_{\psi_1 \wedge \mathcal X (\psi)} \} $. In the case of $U$, we add $UCS:= \{x_\psi \Rightarrow  \mathcal F (x_{\psi_2})\} \cup  UCS$

\item If  $x_\psi \in Set$ is  such that $\psi \in lit(P) \cup \{TRUE;FALSE\}$ then $UCS:=\{x_\psi \Rightarrow \psi \} \cup UCS$



\end{itemize}

with $x_{\psi_j}$ fresh variables at each step. It turns out that $f'=x_f \wedge_{f'' \in UCS} \mathcal G (f'')$ is equi-satisfiable to $f$. Furthermore, $\wedge_{f'' \in UCS} \mathcal G (f'')$ is satisfiable with a model $\mathcal M'$ at $FALSE$ at any time for any propositional variable. This proves the lemma.

\begin{theorem}(\label{MUdecide}MU-dec)\\
Deciding if an unsatisfiable LTL formula is a minimal unsatisfiable formula is  PSPACE-complete.
\end{theorem}
{ \bf (proof)} For any element in  $Weakest-Max(sf^+(f))$, substitute by $TRUE$ and check unsatisfiability. $f$ is a MU iff any substitution leads to a satisfiable formula. There is a linear number of subformulas, and any checking is in $PSPACE$.
  Thanks to lemma \ref{DSNF}  any LTL formula $f$ can be LOGSPACE reduced to an equi-satisfiable formula  of the form  $f'= x_{f} \wedge_{2 \leq i \leq m} f' _i$ with any $f' _i$ is one of the following forms : $\mathcal G (x \vee \mathcal F (x')) $, $ \mathcal G (y \vee \mathcal X (y'))$ or $ \mathcal G (w \vee w' \vee (\delta \wedge w'')) $. Furthermore, let $\mathcal M '$ defined as in lemma \ref{DSNF}. Let $\alpha_1,...,\alpha_m$ some subformulas of $f'$ such that $\alpha_1 =x_f$ and   $f_i' = \mathcal G (\alpha_i)$ for $2 \leq i \leq m$.  Let $x_1,...,x_m$ be fresh boolean variables, and $\pi_i=x_1 \vee .. x_{i-1} \vee x_{i+1} \vee..\vee x_{m}$. Let  "\textrhoticity" be defined as follows: \textrhoticity $l = \sim l$ for $l$ a literal, \textrhoticity $\mathcal X (l) = \mathcal X (\sim l)$ and \textrhoticity $\mathcal F (l) = \mathcal G (\sim l)$. Let $ \omega(f')$ the conjunction of the following  subformulas: $(\mathcal X (x_1)\vee \alpha_1 \vee \pi_1) \wedge_{2 \leq i \leq m} \mathcal G (\alpha_i \vee \pi_i)$, $(\neg x_1 \vee \neg x_f \vee \pi_1) \wedge_{2 \leq i \leq m} \wedge_{op \in \alpha_i}$$\mathcal G$(\textrhoticity$op$$ \vee  \pi_i \vee \neg x_i)$ where $op$ is an operand of the LTL-clause $\alpha_i$,  $\wedge_{1\neq j}(\neg x_1 \vee \neg x_j) \wedge_{2\leq i < j \leq m} \mathcal G (\neg x_i \vee \neg x_j)$, $ \mathcal G (\mathcal X (\neg x_1) )\vee \pi_1$,  $ x_1 \vee...\vee  x_m$. Assume  $\lambda(f')= \omega(f')\setminus \{ x_1 \vee...\vee  x_m\}$. We will show that $\omega(f')$ is a MU iff  $f'$ is satisfiable.\\
A- If $\omega(f')$ is MU then $f'$ is satisfiable.\\
If $\omega(f')$ is MU then $\lambda(f')$ is satisfiable. If $\mathcal M$ is a linear model of $\lambda(f')$, there are  two cases:
(1)  Either any $x_i$ is False at any time point $j$ in $\mathcal M (j)$ and $(\mathcal M,0)\vDash\lambda(f')$ iff $(\mathcal M,0)\vDash f'$. Thus, $f'$ is satisfiable.
(2) Either there exists a time point $j$  and some $x_i$ such that $x_i$ is True in $\mathcal M (j)$. In this case we have:
\begin{itemize}
\item  If $\pi_1$ is FALSE at $\mathcal M (0)$, then thanks to $ \mathcal G (\mathcal X (\neg x_1) )\vee \pi_1$, either $x_1$ is TRUE at $\mathcal M (0)$ and it will never hold later (but in this case $(\mathcal M,0)\nvDash\lambda(f')$ because $(\mathcal M,0)\nvDash (\mathcal X (x_1)\vee \alpha_1 \vee \pi_1) \wedge (\neg x_1 \vee \neg x_f \vee \pi_1)\wedge (\mathcal G (\mathcal X (\neg x_1) )\vee \pi_1) \wedge_{1\neq j'}(\neg x_1 \vee \neg x_{j'})$), or $x_1$ will never hold on $\mathcal M $, then  $i>1$ and $j>0$, but in this  case $(\mathcal M,0)\nvDash \wedge_{2\leq i' < j' \leq m} \mathcal G (\neg x_{i'} \vee \neg x_{j'}) \wedge \mathcal G (\alpha_i \vee \pi_i)\wedge_{op \in \alpha_i}$$\mathcal G$(\textrhoticity$op$$ \vee  \pi_i \vee \neg x_i) \wedge \mathcal G (\mathcal X (\neg x_1) )\vee \pi_1 $ since $(\mathcal M,j)\nvDash \wedge_{2\leq i' < j' \leq m} (\neg x_{i'} \vee \neg x_{j'}) \wedge (\alpha_i \vee \pi_i)\wedge_{op \in \alpha_i}$(\textrhoticity$op$$ \vee  \pi_i \vee \neg x_i)$. Thus $(\mathcal M,0)\nvDash\lambda(f')$
\item   If $\pi_1$ is TRUE at $\mathcal M (0)$ then $i>1$ and for instance $j=0$. We deduce   $(\mathcal M,0)\nvDash \wedge_{1\neq j'}(\neg x_1 \vee \neg x_{j'}) \wedge_{2\leq i' < j' \leq m} \mathcal G (\neg x_{i'} \vee \neg x_{j'})\wedge \mathcal G (\alpha_i \vee \pi_i)\wedge_{op \in \alpha_i}$$\mathcal G$(\textrhoticity$op$$ \vee  \pi_i \vee \neg x_i)  $. Thus $(\mathcal M,0)\nvDash\lambda(f')$.
\end{itemize}
Thus only the case (1) is possible, i.e., $f'$ is satisfiable.\\
B- if  $f'$ is satisfiable  then $\omega(f')$ is MU.\\
Assume $f'$  is satisfiable. We have to show that $\omega (f')$ is (a) unsatisfiable  and (b) minimal.
(1) Assume $\omega (f')$ is satisfiable by a model $\mathcal M$, then only one of $x_i$ is TRUE at $\mathcal M (0)$ but this is the unsatisfiable case  of A-(2), which is a contradiction. Thus  $\omega (f')$ is not satisfiable.
(2) Let $g$ be a subformula in the conjunction of $f'$. We will show that $\gamma (f) = \omega (f') \setminus \{g\}$ is satisfiable for any $g$:\\
- case $g=\mathcal X (x_1)\vee \alpha_1 \vee \pi_1$. Let $\mathcal M$ be a model with  $x_1$ TRUE only at  $\mathcal M(0)$ and FALSE later on and the other $x_{i}$s are always False on $\mathcal M$. Then  $(\mathcal M,0)\vDash\gamma(f')$ iff $(\mathcal M,0)\vDash (\neg x_1 \vee \neg x_f \vee \pi_1)  \wedge_{2 \leq i \leq m} \mathcal G (\alpha_i \vee \pi_i)$, and fixing the other variables as in $\mathcal M '$ is sufficient to show  $(\mathcal M,0)\vDash \gamma (f')$.\\
- case $g=\mathcal G (\alpha_i \vee \pi_i)$. Let $\mathcal M$ be a model with  $x_i$ always TRUE on $\mathcal M $ and  the other $x_{i'}$s are always False on $\mathcal M$, then  $(\mathcal M,0)\vDash\gamma(f')$ iff $(\mathcal M,0)\vDash \wedge_{op \in \alpha_i}$$\mathcal G$(\textrhoticity$op$$ \vee  \pi_i \vee \neg x_i)$. The $op$s does not have the same propositional variables, thus by setting in $\mathcal M $ the literals of $Lit(op)$ always to FALSE for any $op$s of $\alpha_i$  leads to  $(\mathcal M ,0)\vDash\gamma(f')$.\\
- case $g=x_1 \vee...\vee  x_m$ then if any $x_i$ is always FALSE, and since $f'$ is satisfiable and according to 1.(a), $(\mathcal M,0)\vDash\gamma(f')$.\\
- case $g=(\neg x_1 \vee \neg x_f \vee \pi_1) $. Let $\mathcal M$ be a model with  $x_1$ TRUE only at  $\mathcal M(0)$ and FALSE later on and the other $x_{is}$ are always False. Setting the other variables as in a model of $f'$ is sufficient to show  $(\mathcal M,0)\vDash \gamma (f')$.\\
- case $g=\mathcal G$(\textrhoticity$op$$ \vee  \pi_i \vee \neg x_i)$. Let $\mathcal M$ be a model with  $x_i$  always TRUE  and the other $x_{js}$ are always False on $\mathcal M$. Setting literals in  $Lit(op)$ always to FALSE while \textrhoticity$op$ is in $\gamma(f')$ and the  remaining $Lit(op)$ of $\alpha_i$  always at TRUE is sufficient to show  $(\mathcal M,0)\vDash \gamma (f')$.\\
- case $g=\mathcal G (\mathcal X (\neg x_1) )\vee \pi_1$. Let $\mathcal M$ be a model with  $x_1$ always TRUE and the other $x_{is}$ are always False on $\mathcal M$. Setting $x_f$ at FALSE at  $\mathcal M (0)$ is sufficient to show $(\mathcal M,0)\vDash \gamma (f')$.\\
- case $g =\neg x_1 \vee \neg x_j$. Let $\mathcal M$ be a model with  $x_1$ and $x_j$ always TRUE and the other $x_{is}$ are always False on $\mathcal M$. One gets $(\mathcal M,0)\vDash \gamma (f')$.\\
- case $g =\mathcal G (\neg x_i \vee \neg x_j)$. Let $\mathcal M$ be a model with  $x_i$ and $x_j$ always TRUE and the others $x_i$ are always False on $\mathcal M$.  One gets $(\mathcal M,0)\vDash \gamma (f')$.\\
Since the satisfiability decision problem of LTL is  PSPACE complete \cite{SistlaC85}, and it is LOGSPACE reducible to the MU decision problem, MU-decision is PSPACE hard. Consider now the problem of deciding whether, given a LTL formula $g$, there is a strengthening\footnote{some substitutions by $FALSE$ (resp. $TRUE$) of some subformula occurrences of $g$ with positive polarity (resp. negative polarity)} of $g$ which is still equivalent to $g$ (Inherent vacuity with single occurrence \cite{FismanKSV08}).  Let $g = \neg \omega(f)$. Since a PSPACE-complete problem also gets its Co-problem be  PSPACE-complete, one gets:  
\begin{corollary} The inherent vacuity decision problem (with single occurrence) is PSPACE-complete.
\end{corollary}

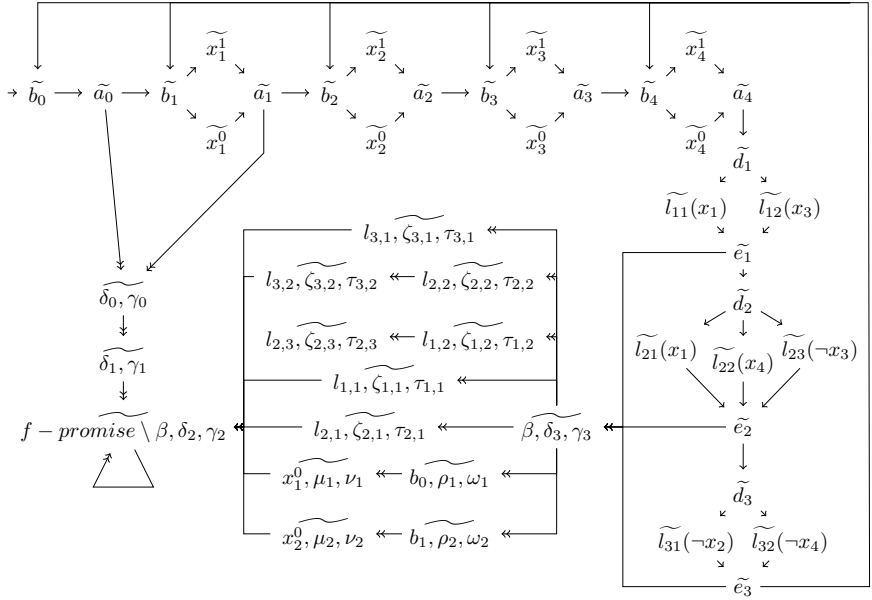
\begin{figure}
\scalebox{0.8}{
\begin{tikzpicture}[shorten >=2pt,node distance=1.1cm,auto]



\node(b_0)  {$\widetilde{b_0}$};
\node(a_0)  [ right of= b_0]{$\widetilde{a_0}$};

\node(b_1) [ right of= a_0] {$\widetilde{b_1}$};
\node(x0_1) [ below right of= b_1] {$\widetilde{x^0 _1}$};
\node(x1_1) [ above right of= b_1] {$\widetilde{x^1 _1}$};
\node(a_1) [ below right of= x1_1] {$\widetilde{a_1}$};
\node(b_2) [ right of= a_1] {$\widetilde{b_2}$};
\node(x0_2) [ below right of= b_2] {$\widetilde{x^0 _2}$};
\node(x1_2) [ above right of= b_2] {$\widetilde{x^1 _2}$};
\node(a_2) [ below right of= x1_2] {$\widetilde{a_2}$};
\node(b_3) [ right of= a_2] {$\widetilde{b_3}$};
\node(x0_3) [ below right of= b_3] {$\widetilde{x^0 _3}$};
\node(x1_3) [ above right of= b_3] {$\widetilde{x^1 _3}$};
\node(a_3) [ below right of= x1_3] {$\widetilde{a_3}$};
\node(b_4) [ right of= a_3] {$\widetilde{b_4}$};
\node(x0_4) [ below right of= b_4] {$\widetilde{x^0 _4}$};
\node(x1_4) [ above right of= b_4] {$\widetilde{x^1 _4}$};
\node(a_4) [ below right of= x1_4] {$\widetilde{a_4}$};

\node(d_1) [ below  of= a_4] {$\widetilde{d_1}$};
\node(l11) [ below left of= d_1, ] {$\widetilde{l _{11}}(x_1)$};
\node(l12) [ below right of= d_1] {$\widetilde{l _{12}}(x_3)$};
\node(e_1) [ below left of= l12] {$\widetilde{e_1}$};
\node(d_2) [ below  of= e_1, yshift=0.3 cm] {$\widetilde{d_2}$};
\node(l21) [ below left of= d_2,xshift=-0.5cm] {$\widetilde{l _{21}}(x_1)$};
\node(l22) [ below  of= d_2,yshift=0.1cm] {$\widetilde{l _{22}}(x_4)$};
\node(l23) [ below right of= d_2,xshift=0.5cm] {$\widetilde{l _{23}}(\neg x_3)$};
\node(e_2) [ below  of= l22] {$\widetilde{e_2}$};
\node(d_3) [ below  of= e_2] {$\widetilde{d_3}$};
\node(l31) [ below left of= d_3] {$\widetilde{l _{31}}(\neg x_2)$};
\node(l32) [ below  right of= d_3] {$\widetilde{l _{32}}(\neg x_4)$};
\node(e_3) [ below left of= l32] {$\widetilde{e_3}$};
\node(beta) [  left  of= e_2,xshift=-2cm] {$\widetilde{\beta, \delta_3, \gamma_3}$};
\node(x_1mu) [   below left  of= beta, xshift=-1cm] {$\widetilde{b_0, \rho_1, \omega_1}$};
\node(b_0rho) [   left  of= x_1mu, xshift=-1cm] {$\widetilde{x^0 _1, \mu_1, \nu_1}$};
\node(x_2mu) [    below left  of= beta , yshift=-1cm,xshift= -1cm] {$\widetilde{b_1, \rho_2, \omega_2}$};
\node(b_1rho) [   left  of=x_2mu,  xshift=-1cm] {$\widetilde{x^0 _2, \mu_2, \nu_2}$};
\node(l_21zeta) [   left  of=beta,  xshift=-2cm] {$\widetilde{l_{2,1}, \zeta_{2,1}, \tau_{2,1}}$};
\node(l_11zeta) [   above left  of=beta,  xshift=-2cm] {$\widetilde{l_{1,1}, \zeta_{1,1}, \tau_{1,1}}$};
\node(l_31zeta) [   above left  of=beta,  xshift=-1.5cm, yshift=2.5cm] {$\widetilde{l_{3,1}, \zeta_{3,1}, \tau_{3,1}}$};
\node(l_12zeta) [   left  of=beta,xshift= -0.2cm, yshift = 1.5cm ] {$\widetilde{l_{1,2}, \zeta_{1,2}, \tau_{1,2}}$};
\node(l_23zeta) [   left  of=l_12zeta,  xshift=-1.5cm ] {$\widetilde{l_{2,3}, \zeta_{2,3}, \tau_{2,3}}$};
\node(l_22zeta) [   left  of=beta,  xshift=-0.2cm, yshift = 2.5cm ] {$\widetilde{l_{2,2}, \zeta_{2,2}, \tau_{2,2}}$};
\node(l_32zeta) [   left  of=l_22zeta,  xshift=-1.5cm ] {$\widetilde{l_{3,2}, \zeta_{3,2}, \tau_{3,2}}$};
\node(promiseminusbeta) [  left  of= l_21zeta,xshift=-3cm] {$\widetilde{f-promise \setminus \beta, \delta_2, \gamma_2}$};
\node(delta1) [  above  of= promiseminusbeta] {$\widetilde{ \delta_1, \gamma_1}$};
\node(delta0) [  above  of= delta1] {$\widetilde{ \delta_0, \gamma_0}$};
\draw[->] +(-0.5,0)--(b_0);

\draw[->>](promiseminusbeta) -- +(0.5,-1)-- +(-0.5,-1)-- (promiseminusbeta);
 \draw[->] (e_3) -- +(2.1,0) -- (13.8,1.5) -| (b_0) ;
 \draw[->] (13.5,1.5) -| (b_1);
 \draw[->] (13.5,1.5) -| (b_2);
 \draw[->] (13.5,1.5) -| (b_3);
 \draw[->] (13.5,1.5) -| (b_4);
\draw[->>, ] (a_0) --  (delta0);
\draw[->>, ] (a_1) --+(0,-1)--  (delta0);
\draw[->>, ] (delta0) --  (delta1);
\draw[->] (b_0) --  (a_0);
\draw[->] (a_0) --  (b_1);
\draw[->] (b_1) --  (x0_1);
\draw[->] (b_1) --  (x1_1);
\draw[->]   (x0_1)-- (a_1);
\draw[->]   (x1_1)-- (a_1);
\draw[->] (b_2) --  (x0_2);
\draw[->] (b_2) --  (x1_2);
\draw[->]   (x0_2)-- (a_2);
\draw[->]   (x1_2)-- (a_2);
\draw[->] (a_1) --  (b_2);
\draw[->] (a_2) --  (b_3);
\draw[->>, ] (l_22zeta) --  (l_32zeta);
\draw[->>, ] (l_12zeta) --  (l_23zeta);
\draw[->] (a_3) --  (b_4);
\draw[->] (b_3) --  (x0_3);
\draw[->] (b_3) --  (x1_3);
\draw[->]   (x0_3)-- (a_3);
\draw[->]   (x1_3)-- (a_3);
\draw[->] (b_4) --  (x0_4);
\draw[->] (b_4) --  (x1_4);
\draw[->]   (x0_4)-- (a_4);

\draw[->]   (x1_4)-- (a_4);
\draw[->>, ] (delta1) -- (promiseminusbeta);
\draw[->] (d_1) -- (l12);
\draw[->]  (l12)-- (e_1);
\draw[->] (d_1) -- (l11);
\draw[->]  (l11)-- (e_1);
\draw[->] (d_2) -- (l21);
\draw[->]  (l21)-- (e_2);
\draw[->] (d_2) -- (l22);
\draw[->]  (l22)-- (e_2);
\draw[->] (d_2) -- (l23);
\draw[->]  (l23)-- (e_2);
\draw[->] (d_3) -- (l31);
\draw[->]  (l31)-- (e_3);
\draw[->] (d_3) -- (l32);
\draw[->]  (l32)-- (e_3);
\draw[->] (a_4) --  (d_1);
\draw[->] (e_1) --  (d_2);
\draw[->] (e_2) --  (d_3);
\draw[->>, ] (beta) |- (l_22zeta);
\draw[->>, ] (beta) |- (l_31zeta);
\draw[->>, ] (beta) |- (l_12zeta);
\draw[->>, ] (beta) |- (l_11zeta);
\draw[->>, ] (beta) -- (l_21zeta);
\draw[->>, ] (beta) |- (x_1mu);
\draw[->>, ] (x_1mu) --  (b_0rho);
\draw[->>, ] (x_2mu) --  (b_1rho);
\draw[->>, ] (beta) |- (x_2mu);
\draw[->>, ] (e_3) -- +(-2,0)|- (beta);
\draw[->>, ] (e_2) -- +(-2,0)|- (beta);
\draw[->>, ] (e_1) -- +(-2,0)|- (beta);
\draw[<<-, ]  (promiseminusbeta) -- +(2,0) |- (l_11zeta);
\draw[<<-, ]  (promiseminusbeta) -- +(2,0) |- (l_31zeta);
\draw[<<-, ] (promiseminusbeta) -- +(2,0) |- (l_21zeta);
\draw[<<-, ]  (promiseminusbeta) -- +(2,0) |- (l_32zeta);
\draw[<<-, ] (promiseminusbeta) -- +(2,0) |- (b_0rho);
\draw[<<-, ]  (promiseminusbeta) -- +(2,0) |- (b_1rho);
\end{tikzpicture}}
\caption{$\mathcal K $ and $\mathcal K '$ for $\Phi=\forall x_1 \forall x_2 \exists x_3 \exists x_4 (x_1 \vee x_3) \wedge (x_1 \vee x_4 \vee \neg x_3) \wedge (\neg x_2 \vee \neg x_4)$ \label{dfs}}
\end{figure}
{\bf Canonical reduction of QCNF into LTL Model Checking \cite{DemriS98}}\\
 
 Let $\Phi$ be a closed  $QCNF$. Thus, $\Phi=Q \overline x \phi$ with $\phi \in CNF$  is of the form $\phi =\wedge_{1 \leq j \leq m} C_j$, where any $C_j$ is a clause   and $|\overline x| = n$. We begin to recall the existence of a Kripke structure $\mathcal K$ and a LTL formula 
$\Psi$ such that: { \bf$ {\mathcal{ K}\vDash \Psi} $ iff $\mathbf{\Phi}$ is FALSE} \cite{DemriS98}.\\
We start by  the example $\Phi=\forall x_1 \forall x_2 \exists x_3 \exists x_4 (x_1 \vee x_3) \wedge (x_1 \vee x_4 \vee \neg x_3) \wedge (\neg x_2 \vee \neg x_4)$. The  Kripke structure $\mathcal K$ is shown  Figure \ref{dfs}.  For space commodity, $\mathcal K$  is indicated by the arrows with simple arrowhead (do not consider double arrowheads). Intuitively  a path at the `above' part of $\mathcal K$ instantiates the variables of $\phi$ (by choosing to display $x^0 _i$ or $x^1 _i$), and  a path at the `right' part displays any choosen literal per clause ($l_{j,k}$). Consistency constraints (1) between instances of variables and displayed literal per clause and (2) to express universal quantifier of $\Phi$ are expressed in the LTL formula $\Psi$. In the general case $\mathcal K $ and $ \Psi$ are defined as follows:

 Let  $P$ be the set of the following fresh propositional variables for LTL formula: 
\begin{multicols}{2}
\begin{itemize}
\item $a_0$, $b_0$
\item $\forall i$ $1 \leq i \leq n$, $x^0 _i$, $x^1 _i$, $a_i$ , $b_i$
\item $\forall j$ $1 \leq j \leq m$, $d_j$, $e_j$, $\forall k$, $l_{(j,k)}$ with $C_j=\vee_{ 1\leq k \leq k_j} lit(l_{(j,k)})$ 
\end{itemize}  
\end{multicols}
Moreover, any $l_{(j,k)}$ or $x^\epsilon_i$ is a propositional variable standing for a literal written down $lit(l_{(j,k)})$ as a literal of $C_j$, $lit(x^0_i)= \neg x_i$ and $lit(x^1_i)=   x_i$, with $x_i \in  \overline x$.\\
Let a Kripke structure $\mathcal K=(S,\{\widetilde {b_0}\},T,l)$ where   any state is defined by its label. For $s \in S$, if $l(s)= \{p_1,..., p_q\}$, then  $ \widetilde{(p_1,..., p_q)}$ denotes $s$. Furthermore,  $\widetilde{b_0}$ is the  solely starting state. $\mathcal K$ is the smallest Kripke structure allowing the following transitions: 
\begin{multicols}{2}
\begin{itemize}
\item $(\widetilde {b_0}, \widetilde {a_0})$
\item  $(\widetilde{a_i},\widetilde{b_{i+1}})$ for any $ i$, $0 \leq i\leq n-1$
\item  $(\widetilde{b_i},\widetilde{x^0 _i})$ , $(\widetilde{b_i},\widetilde{x^1 _i})$, $(\widetilde{x^0 _i},\widetilde{a_{i}})$, $(\widetilde{x^1 _i},\widetilde{a_{i}})$ for any $ i$, $1 \leq i\leq n$
\item $(\widetilde{a_n}, \widetilde{d_1})$
\item $(\widetilde{d_j},\widetilde{l_{(j,k)}})$ , $(\widetilde{l_{(j,k)}}, \widetilde{e_{j}})$ for any $ j,k$ such that  $1 \leq j \leq m$, $ 1\leq k \leq k_j$,
\item  $(\widetilde{e_j},\widetilde{d_{j+1}})$ for any $j |1 \leq j \leq m-1$ 
\item  $(\widetilde{e_m},\widetilde{b_{i}})$ for any  $ i |0 \leq i \leq n$ ,
\end{itemize}
\end{multicols}

 In case of no confusion, we  will write $\widetilde {p_1,...,p_q}$ to denote the canonical conjunction of its literals: $\wedge_{1 \leq t \leq q} p_t  \wedge_{p \in P\setminus (\cup_{1 \leq t \leq q} \{ p_t \} )} \neg p $ .
Let $\Psi_{univ}=\wedge_{i|Q_i=\forall} \mathcal G (\widetilde {a_{i-1}} \Rightarrow [((\neg b_{i-1})\, \mathcal U x^0 _i) \wedge \mathcal{X}^2(x^1_i)])$ standing for the constraints enforcing  a potential $\mathcal K$ linear model  to first visit the state $\widetilde {x^1 _i}$ for any $Q_i = \forall$ and later visit $\widetilde {x^0 _i}$. Furthermore, as long as both states have not  been visited, any  $\mathcal K$ linear model cannot go back to the previous $\widetilde {x^\epsilon _j}$ states for $j<i$. Let $\Psi_{cons} =\wedge_{(i,j,k,\epsilon) | lit(x^{\epsilon}_{i}) = \sim lit(l_{(j,k)}) } \mathcal G (\widetilde{x^{\epsilon}_{i}} \Rightarrow ((\neg l_{(j,k)})\mathcal W b_i))$  which stands for the constraints enforcing the consistency of the instantiated variables $x_i$ of $\Phi$ at FALSE ($ x^{0}_{i}$) or at TRUE ($x^{1}_{i}$) with their opposite literal occurring in the clauses. Finally, let $\Psi= \neg (\Psi_{univ} \wedge \Psi_{cons})$. Then, the authors of \cite{DemriS98} have shown that $\mathcal K \vDash \Psi $ iff $\Phi$ is FALSE.\\

{\bf Searching  Minimal Unsatisfiable LTL Formulas }
\begin{theorem}(MU-Search) \label{MUsearch} Given  a LTL $f$, providing a MU of $f$ if $f$ is unsatisfiable, and answer `no' if $f$ is satisfiable is a FPSPACE complete problem.
\end{theorem}

{\bf (proof)}\\ 
Assume the same notations of  the canonical reduction. At this step, encoding $\mathcal K$ into a LTL formula $\Phi_\mathcal K= \widetilde b_0 \wedge_{\widetilde s \in S } \mathcal G (\widetilde s \Rightarrow \mathcal X (\vee_{(\widetilde s,\widetilde { s'}) \in T  } \widetilde {s'})$ and study the extraction of a MU of $\Phi_\mathcal K \wedge \neg \Psi $ is tempting. However, for instance, it cannot ensure that a MU of $\Phi_\mathcal K \wedge \neg \Psi $ gets a corresponding MF because a MU may be without some `universal' subformulas occurrences of the form $sf_i = \mathcal G (\widetilde {a_{i-1}} \Rightarrow [((\neg b_{i-1})\, \mathcal U x^0 _i) \wedge X^2(x^1_i)])$ which may not lead to a QBF (e.g., if $\Phi=\forall a \forall b  (a \vee b) \wedge (\neg a \vee b)$, then $g= [\Phi_\mathcal K \wedge \neg \Psi ][sf_1\leftarrow TRUE]$ is still  unsatisfiable and a MU of $g$ does not correspond to any MF of $\Phi$).
 We then have to create a new Kripke structure $\mathcal K'$  by adding variables and  several branches in $\mathcal K$ (supported by double arrowheads) to  enforce most of the subformulas occurrences to be necessary, i.e., to remain in any MU of  $\Phi_{\mathcal K'} \wedge  \neg \Psi$. To do so we also need to weaken $\neg \Psi$ by adding  disjuncts which are  promises to fulfil $(\mathcal F (\beta))$.  Unfortunately, we also have to tightly  strengthen the just resulting weakened new formula (into $\neg \Psi'$) by adding conjuncts, in order that the new branches do not imply the existence of a $\mathcal K'$ linear model of $\neg \Psi'$ while $\Phi $ is FALSE.  Finally, it is still possible to find such $\mathcal K'$ and $\Psi'$ such that $\mathcal{ K}' \vDash \Psi' $ iff $\Phi$ is FALSE (see point A). Then, the proof reduces the latter Model Checking  problem to  a variant of an LTL unsatisfiability problem (see $Temp(\Phi)$ at point B). Finally, the analysis of the $MU(Temp(\Phi))$ (see section C) regarding the $MF(\Phi)$ enables to show the MF search problem can be reduced to the MU search problem by focusing on mutations of the $l_{(j,k)}$ at the state $\widetilde{l_{(j,k)}}$.\\

{\bf \large A-} \normalsize \underline { $\mathcal{ K}' \vDash \Psi' $ iff $\Phi$ is FALSE}\\

Let $P'$ be $P$ augmented with the following variables:   $\beta$; $\forall r$ $0 \leq r \leq 3$, $\delta_r$, $\gamma_r$;  $\forall i$ $1 \leq i \leq n$, $\mu_i$, $\nu_i$, $\omega_i$, $\rho_i$; $\forall j$ $1 \leq j \leq m$,  $\tau_{(j,k)}$ and $\zeta_{(j,k)}$ with $C_j=\vee_{ 1\leq k \leq k_j} lit(l_{(j,k)})$. For convenience, we define $\widetilde {p_1,..,p_q}$ as a state/conjunction with the corresponding literals over $P'$. For technical reasons, from now,  we  similarly will extend a  state $\widetilde {p_1,..,p_q}$ of $\mathcal K$ to $\mathcal K '$ by adding the new (negated) variables to the corresponding conjunction. Let the set $f-promises = \cup_{0 \leq i \leq n} \{a_i,b_i; x^0_i \} \cup \{ \beta  \} $. The resulting  Kripke structure $ \mathcal K'$ adds the following transitions to $\mathcal K$:
\begin{multicols}{2}
\begin{itemize}

\item  $(\widetilde{a_{i-1}},\widetilde{(\delta_0,\gamma_0)})$ for any  $i| Q_i=\forall$
\item  $(\widetilde{(\delta_0,\gamma_0)},\widetilde{(\delta_1,\gamma_1)})$ 
\item$(\widetilde{(\delta_1,\gamma_1)}, \widetilde{(f.promises\setminus \{\beta \},\delta_2,\gamma_2)})$ 
\item $(\widetilde e_j,\widetilde {(\beta,\delta_3,\gamma_3)})$ for any $j$ with $1 \leq j \leq m$
\item  $(\widetilde {(\beta,\delta_3,\gamma_3)}, \widetilde{(b_{i-1},\rho_i,\omega_i)})$ for any $  i$ such that  $i| Q_i=\forall$
\item  $(\widetilde{ (\beta,\delta_3,\gamma_3)}, \widetilde {(l_{(j,k)},\zeta_{(j,k)},\tau_{(j,k)})})$ for any $ j,k$ such that  $1 \leq j \leq m$ and    there is no $j',k'$ such that $lit (l_{(j,k)}) = \sim lit (l_{(j',k')})$,
\item $( \widetilde {(\beta,\delta_3,\gamma_3)},\widetilde {(l_{(j,k)},\zeta_{(j,k)},\tau_{(j,k)})})$ for any $ j, k$  such that $lit (l_{(j,k)})$  is positive,  $1 \leq j \leq m$ and  $\exists j',k'$ such that $ lit (l_{(j,k)}) =\sim lit (l_{(j',k')})$.
\item $(\widetilde{(f.pro.\setminus \{\beta \},\delta_2,\gamma_2)},$ $ \widetilde{(f.pro.\setminus \{\beta \},\delta_2,\gamma_2)})$\\

\item $( \widetilde {(l_{(j,k)},\zeta_{(j,k)},\tau_{(j,k)})},$ $ \widetilde {(l_{(j',k')},\zeta_{(j',k')},\tau_{(j',k')})})$ for any  $ j,j', k,k'$ such that  $1 \leq j,j' \leq m$ and such that $lit (l_{(j,k)}) =\sim lit (l_{(j',k')})$. Furthermore  $lit(l_{(j,k)})$ is a positive literal.
\item  $ (\widetilde{(b_{i-1},\rho_i,\omega_i)},\widetilde{(x_{i}^0,\mu_i,\nu_i)})$ for any  $i| Q_i=\forall$
\item  $ (\widetilde{(x_{i}^0,\mu_i,\nu_i)},$$(\widetilde{f.promises \setminus \{\beta \},\delta_2,\gamma_2}))$ for any  $i| Q_i=\forall$
\item $( \widetilde {(l_{(j,k)},\zeta_{(j,k)},\tau_{(j,k)})},$ $\widetilde {f.promises \setminus \{ \beta \},\delta_2,\gamma_2})$ for any  $ j, k$ such that  $1 \leq j \leq m$ and such there is no $j',k'$ such that $lit (l_{(j,k)}) =\sim lit (l_{(j',k')})$ 
\item $( \widetilde {(l_{(j,k)},\zeta_{(j,k)},\tau_{(j,k)})},$ $\widetilde {f.promises \setminus \{ \beta \},\delta_2,\gamma_2})$ for any  $ j, k$ such that  $1 \leq j \leq m$ ,such $\exists j',k'$ such that $lit (l_{(j,k)}) =\sim lit (l_{(j',k')})$. Furthermore  $l_{(j,k)}$ is a negative literal.

\end{itemize}
\end{multicols}

In Figure \ref{dfs}, $ \mathcal K '$ is  supported by all the arrows (with simple or double arrowhead).
Let $\Psi' = \neg (\Psi_{univ'} \wedge \Psi_{cons'})$ with $\Psi_{univ'}= \wedge_{i|Q_i=\forall} [\mathcal G (\widetilde {a_{i-1}} \Rightarrow [(\neg b_{i-1} \vee \mathcal F (\beta))\, \mathcal U x^0 _i) \wedge (X^2(x^1_i) \vee \mathcal F (\beta))])] \wedge \mathcal G ((\widetilde b_i \wedge \mathcal X (x_i ^ 0))  \Rightarrow X^2((\neg x_i ^0 \vee \mathcal F (\beta))\mathcal W a_{i-1}) $ , and $ \Psi_{cons'}=\wedge_{(i,j,k,\epsilon) | lit(x^{\epsilon}_{i}) = \sim lit(l_{(j,k)}) } \mathcal G (\widetilde{x^{\epsilon}_{i}} \Rightarrow ((\neg l_{(j,k)} \vee  \mathcal F (\beta))\mathcal W b_i)) $\\ $\wedge_{(i,j,k,\epsilon) | lit(x^{\epsilon}_{i}) =  lit(l_{(j,k)})  ,\forall j',k' lit(l_{j',k'}) \neq \sim lit(l_{(j,k)}) } \mathcal G (\widetilde{x^{\epsilon}_{i}} \Rightarrow ((\neg \zeta_{(j,k)} \vee \neg \tau_{(j,k)} \vee \mathcal F (\beta))\mathcal W b_i))$.\\

In the following, we show that a $\mathcal {K'}$-linear model of $\neg \Psi'$ is necessarily a $\mathcal {K}$-linear model of $ \neg \Psi'$. This  implies\footnote{From a $\mathcal K$-linear model of $ \neg \Psi$ we can derive a $\mathcal K$-linear model of $\neg \Psi \wedge \mathcal G ((\widetilde b_i \wedge \mathcal X (x_i ^ 0))  \Rightarrow X^2((\neg x_i ^0 \vee \mathcal F (\beta))\mathcal W a_{i-1}) $, which is also a $\mathcal K$-linear model of $\neg \Psi'$ } $\mathcal K' \vDash \Psi' $ iff $\Phi$ is FALSE:
\begin{itemize}
\item $\widetilde {(l_{(j,k)},\zeta_{(j,k)},\tau_{(j,k)})}$  cannot occur in a  $\mathcal K'$- linear model of $\neg \Psi'$ because for such a model   $\forall  i \in  [1;n]$, there exists $\epsilon$,   such that $\widetilde x^{\epsilon}_{i}$ is the last occurrence of a $\widetilde x^{\epsilon'}_{i}$ before  the  first visit of $\widetilde {(l_{(j,k)},\zeta_{(j,k)},\tau_{(j,k)})}$. Furthermore,
\begin{itemize}
\item if there are no $j',k'$ such that $lit(l_{(j,k)}) = \sim lit(l_{(j',k')}) $  then there is a $\mathcal G (\widetilde{x^{\epsilon}_{i}} \Rightarrow (( g \vee \mathcal F (\beta)) \mathcal W b_i))$ with $g \in \{\neg \zeta_{(j,k)}\vee \neg \tau_{(j,k)} ; \neg l_{(j,k)} \}$ occurring in $\neg \Psi'$ such that the weak promise $( g \vee \mathcal F (\beta)) \mathcal W b_i$ is postponed from $\widetilde x^{\epsilon}_{i}$ to  $\widetilde {(l_{(j,k)},\zeta_{(j,k)},\tau_{(j,k)})}$, but at this latter state $\mathcal F (\beta)$ must hold, which is impossible.
\item  if there exist $j',k'$ such that $lit(l_{(j,k)}) = \sim lit(l_{(j',k')}) $ then  there exists   $\mathcal G (\widetilde{x^{\epsilon}_{i}} \Rightarrow (( g \vee \mathcal F (\beta)) \mathcal W b_i))$ occurring in $\neg \Psi'$ with $g \in \{\neg  l_{(j',k')}, \neg l_{(j,k)} \}$ such that the weak promise  $(g \vee \mathcal F (\beta)) \mathcal W b_i$ is postponed from $x_i^\epsilon$ until  the corresponding $\widetilde {(l_{(j',k')},\zeta_{(j',k')},\tau_{(j',k')})}$ or $\widetilde {(l_{(j,k)},\zeta_{(j,k)},\tau_{(j,k)})}$, but at this convenient latter state $\mathcal F (\beta)$ must hold, which is impossible.
\end{itemize}
\item   $(\widetilde{b_{i-1},\rho_i,\omega_i})$   cannot occur in a  $\mathcal K'$ linear model  of $\neg \Psi'$. Assume $\mathcal M$ such a model.
Let $t$ be the last time where $\widetilde a_{i-1}$ occurs in $\mathcal M$. Then, either:
\begin{itemize}
\item $\widetilde{ x_i ^0}$ occurs in $\mathcal M _t$, but thanks to $\mathcal G ((\widetilde b_i \wedge \mathcal X (x_i ^ 0))  \Rightarrow X^2((\neg x_i ^0 \vee \mathcal F (\beta))\mathcal W a_{i-1}) $, $(\neg x_i ^0 \vee \mathcal F (\beta))\mathcal W a_{i-1}$ is postponed from $\widetilde x_i ^0$ to $\widetilde{ x_i ^ 0 ,\mu_i ,\nu_i  }$. But at this latter state $\mathcal F (\beta)$ must hold, which is impossible.
\item or $\widetilde {x_i ^0}$ does not occur in $\mathcal M _t$, but thanks to $\mathcal G (\widetilde {a_{i-1}} \Rightarrow (\neg b_{i-1} \vee \mathcal F (\beta))\, \mathcal U x^0 _i))$, $(\neg b_{i-1} \vee \mathcal F (\beta))\, \mathcal U x^0 _i$ is postponed from $\widetilde a_{i-1} $ to $\widetilde{ b_{i-1} ,\rho_i ,\omega_i  }$. But at this latter state $\mathcal F (\beta)$ must hold, which is impossible.
\end{itemize}
\item  $\widetilde{(\delta_1,\gamma_1)}$ cannot occur in a   $\mathcal K'$ linear model of $\neg \Psi'$  because for such a model, $\mathcal G (\widetilde {a_{i-1}} \Rightarrow [ X^2(x^1_i) \vee \mathcal F (\beta)])$  implies   $\mathcal F (\beta)$ is propagated from $\widetilde {a_{i-1}}$ to $\widetilde{(\delta_1,\gamma_1)}$, which is impossible.

\end{itemize}

Below, we define $\Psi_{\mathcal K '}$ which stands for $\mathcal K'$. At the next step of the proof, it will enable to reduce the MF search problem to a MU search problem for LTL. To do so, we denote $Temp(\Phi)=\Psi_\mathcal {K'} \wedge \neg \Psi'$ with $\Psi_\mathcal {K'} $ defined in the following. It is then straightforward that $Temp(\Phi)$ is
 unsatisfiable iff $\Phi$ is FALSE. \\ 

\underline{{\bf \large B- \normalsize} { \bf $Temp(\Phi)=\Psi_\mathcal {K'} \wedge \neg \Psi'$ is unsatisfiable iff $\Phi$ is FALSE}}\\
 Let $\Psi_{\mathcal K'}$  as $\Phi_{\mathcal K '}$  except that the occurrences $\mathcal G(s \Rightarrow X(...))  $ where   $s= \widetilde {l_{j,k}} $  are erased. Furthermore one adds the conjuncts $ \mathcal G (\widetilde{d_j} \Rightarrow X^2 \widetilde{e_{j}})$ for any $j|1 < j \leq m$.

We have $Temp(\Phi)=\Psi_\mathcal {K'} \wedge \neg \Psi'$ is unsatisfiable iff $\Phi$ is FALSE.\\

In the following, we analyze that an element of $MU(Temp(\Phi))$  corresponds to some  maximal mutations of propositional variables $l_{(j,k)}$
 at the corresponding states $\widetilde {l_{(j,k)}}$ in $\mathcal K'$ but which the resulting mutated Kripke structure still checks $\Psi_{\mathcal K'}$.  This enables to show that there exists a corresponding  element in $MF(\Phi)$. \\

{\bf \large C- \normalsize}\underline {  Analysis of a $MU(Temp(\Phi))$} \\

Let $MU_0(Temp(\Phi))\in  MU(Temp(\Phi))$. 
\begin{enumerate}
\item the universal part $\Psi _{univ'}$ still occurs in $MU_0(Temp(\Phi))$: 
\begin{itemize}
\item Let  $i$ be an integer such that $Q_i=\forall$ and assume $\neg b_{i-1} \vee \mathcal F (\beta)$ has been substituted by $TRUE$  while  weakening from $Temp(\Phi)$ to  $MU(Temp(\Phi))$ at the universal part. Let $\mathcal M$ be a $\mathcal K'$  linear structure  starting with the state $\widetilde {b_0}$, which never visits $\widetilde {x_i ^0}$ and  reaches $\widetilde{(b_{i-1},\rho_i,\omega_i)}$. From $\widetilde {(\beta, \delta_3,\gamma_3)} $, no constraint enables to propagate $\mathcal F (\beta)$, and any other triggered and postponed  promises  are  fulfilled at  $(\widetilde{f-promises \setminus \{\beta \},\delta_2,\gamma_2})$. From $(\widetilde{f-promises \setminus \{\beta \},\delta_2,\gamma_2})$ any constraint from $MU_0(Temp(\Phi))$ is obviously checked. Then $\mathcal M$ is a model of $MU_0(Temp(\Phi))$, which is a contradiction.

\item Let  $i$ be an integer such that $Q_i=\forall$ and assume that $\mathcal G (\widetilde {a_{i-1}} \Rightarrow [ X^2(x^1_i )\vee \mathcal F (\beta)])$ has been substituted by $TRUE$. Let $\mathcal M$ be a $\mathcal K'$  linear structure starting with the state $\widetilde {b_0}$, directly reaching  $\widetilde{a_{i-1}}$ but by crossing any $x^1 _{i'}$ with $ 1 \leq i' \leq i-1$  and from  $\widetilde{a_{i-1}}$ follows the solely branch where  $\widetilde{(\delta_1,\gamma_1)}$ occurs. $\mathcal M$  is a linear model for $MU_0(Temp(\Phi))$, because at $\widetilde{(\delta_1,\gamma _1)}$ any postponed  weak promise is solely the remaining subformula  $ (\neg b_{i'-1} \vee \mathcal F (\beta))\, \mathcal U x^0 _{i'}$ or $(g \vee \mathcal F (\beta))\mathcal W b_{i'} $ and they are  fulfilled at the first visit of $ \widetilde{(f-promises \setminus \{\beta\},\delta_2,\gamma_2)}$. From $ \widetilde{(f-promises \setminus \{\beta\} ,\delta_2,\gamma_2)}$, any constraint from $MU_0(Temp(\Phi))$ is obviously checked. Thus $MU_0(Temp(\Phi))$ is satisfiable, but this is a contradiction since it is unsatisfiable. Thus, $MU_0(Temp(\Phi))$  does not get any  weakening of $\mathcal G (\widetilde {a_{i-1}} \Rightarrow [ X^2(x^1_i) \vee \mathcal F (\beta)])$ for any $i | Q_i= \forall$. 

\item Let $i$ be an integer such that $Q_i=\forall$ and assume that $\mathcal G ((\widetilde {b_{i-1}} \wedge \mathcal X (x_i ^ 0)) \Rightarrow  X^2[(\neg x^0_i \vee \mathcal F (\beta))\mathcal W a_{i-1}])$ has been substituted by $TRUE$. Let $\mathcal M$ be a $\mathcal K'$  linear structure starting with the state $\widetilde {b_0}$,  reaching $\widetilde{x^0 _{i}}$, and going straightforward to the solely branch where  $\widetilde{(x_{i} ^0 ,\mu_i,\nu_i)}$ occurs. $\mathcal M$  is a linear model for $MU_0(Temp(\Phi))$, because from $\widetilde {(\beta, \delta_3,\gamma_3)}$, no constraint enables to propagate $\mathcal F (\beta)$, and any other remaining precedent propagated promises  are  fulfilled at  $(\widetilde{f-promises \setminus \{\beta \},\delta_2,\gamma_2})$. From $(\widetilde{f-promises \setminus \{\beta \},\delta_2,\gamma_2})$ any constraint is obviously checked. Then $\mathcal M$ is a model of $MU_0(Temp(\Phi))$, which is a contradiction.
\end{itemize}

\item the consistency part $\Psi _{cons'}$ still occurs in $MU_0(Temp(\Phi))$: 
\begin{itemize}
\item Assume that $\mathcal G (\widetilde{x^{\epsilon}_{i}} \Rightarrow (( g \vee \mathcal F (\beta))\mathcal W b_i))$ with $g \in \{\neg \zeta_{(j,k)}\vee \neg \tau_{(j,k)} ; \neg l_{(j,k)} \}$ has been substituted by $TRUE$  while  weakening from $Temp(\Phi)$ to  $MU(Temp(\Phi)$ at the consistency part. Let $\mathcal M$ be a $\mathcal K'$  linear structure starting with the state $\widetilde {b_0}$, going through $\widetilde{x^{\epsilon}_{i}}$  and directly reaching   $\widetilde{e_{1}}$ and from  $\widetilde{e_{1}}$, it  follows a branch where $\widetilde {(l_{(j,k)},\zeta_{(j,k)},\tau_{(j,k)})}$  occurs. From $\widetilde {(\beta, \delta_3,\gamma_3)}$, no constraint enables to propagate $\mathcal F (\beta)$. This is because  no weak promise\footnote{i has been fixed here} of the form $(\neg g'\vee \mathcal F (\beta))\mathcal Wb_i$ must hold anymore. As usual, any other remaining propagated promises  are  fulfilled at  $(\widetilde{f-promises \setminus \{\beta \},\delta_2,\gamma_2})$. Thus $MU_0(Temp(\Phi))$ is satisfiable, but this is a contradiction since it is unsatisfiable.

\end{itemize}

\item   $MU_0(Temp(\Phi))$ gets no weakening in $\Psi_\mathcal {K'} \setminus [\cup_{1 \leq j \leq m} Weakestmax_{sf^+}(\mathcal G (\widetilde{d_j} \Rightarrow X (\vee_{ 1\leq k \leq k_j}\widetilde{l_{(j,k)}}))) ]$.\\
 Otherwise if  there is any weakening in the computation of $MU_0(Temp(\Phi))$ from\\ $\Psi_\mathcal {K'} \setminus \cup_{1 \leq j \leq m} Weakestmax_{sf^+}(\mathcal G (\widetilde{d_j} \Rightarrow X (\vee_{ 1\leq k \leq k_j}\widetilde{l_{(j,k)}}))) $, then $MU_0(Temp(\Phi))$ would be satisfiable because of the following points.
 \begin{itemize}
\item any weakening in  a subformula occurrence $\mathcal G ( s \Rightarrow \mathcal X^k (\vee_{( s, s') \in T  } s') $ of  $Temp(\Phi)$ which is  not of the form  $\mathcal G (\widetilde{d_j} \Rightarrow X (\vee_{ 1\leq k \leq k_j}\widetilde{l_{(j,k)}}))$ leads to at least a weakening of a literal $l \in lit^+$ at $s'$ which may lead at a state which is not in $\mathcal K'$. To show this property, consider the $ s $ at the weakened above subformula occurrence and let $pref$ a prefix linear structure in $\mathcal K'$ such that its last state is either (1) the  $ s \in \mathcal K' $ if $k=1$ either (2) any $l_{j,k}$ while $ s = \widetilde {d_j}$ and $k=2$.  Let $s_0$ and $s_1$ some states over $P'$, we define $|s_0 - s_1 | = |\{p | p \in( l(s_0) \setminus l(s_1)) \cup (l(s_1) \setminus l(s_0))\}|$. It is clear that for any $ s_0$ and $ s_1$ in $\mathcal K' $, $|s_0 - s_1 |>1$. Let $s''$ such that $s''=s'[l \leftarrow  \sim l]$. It is clear that $|s' - s''|=  1$. It turns out that $s''\notin \mathcal K'$. Let $\mathcal M$  the linear structure $\mathcal M = pref. s''.\widetilde {P'} ^\omega$. From the state  $s'' $, $MU_0(Temp(\Phi)) \cap \neg \Psi_\mathcal {K'}$ is trivially checked since $s'' \notin \mathcal K'$ and $\widetilde {P'} \notin \mathcal {K'}$. Furthermore, at the first visit of the  state  $\widetilde P'$   the weak promises postponed and propagated from $s''$ are fulfilled, and from  $\widetilde {P'}$  any propagated temporal constraint from $MU_0(Temp(\Phi))$ is obviously checked because $ \widetilde {P'} \notin \mathcal K'$. Thus, $\mathcal M \vDash MU_0(Temp(\Phi))$.

\item For similar argument any weakening in $s=\widetilde b_0 $ ensures that a linear model of $ MU_0(Temp(\Phi))$ is $(s')^\omega$ with $s' \notin \mathcal {K'} $.
\end{itemize}

\item Let $ H \subset [1,m]$ and $S_f= \cup_{j \in H} \{f_j\}$ the  subformulas\footnote{ One formula per $j$ is sufficient } of the form $f_j=\mathcal G (\widetilde{d_j} \Rightarrow X (\vee_{ 1\leq k \leq k_j}\widetilde{l_{(j,k)}})) $ for any $j|1 < j \leq m$  which have been  weakened in $Temp(\Phi)$ by TRUE at a positive occurrence of the  variable $l_{(j,k)}$ in $\widetilde {l_{(j,k)}}$ for computing $MU_0(Temp(\Phi))$. Then $\Phi[C_h\leftarrow TRUE]_{h \in H}$ is in $MF(\Phi)$.\\
 The reasons are summarized below.

\begin{itemize}

\item $\neg \beta$ is not weakened at any $\widetilde{l_{(j,k)}}$ in any $f_j$? Let $s_0 = \widetilde{(l_{(j,k)} , \beta)}$. Otherwise, a linear structure of $\mathcal K \cup \{s_0\}$ which checks the rules of $\neg \Psi_{univ'}$ and sometimes which visit the  `weakened' state $s_0$ where $\beta \in l(s_0)$ is a model of $MU_0(Temp(\Phi))$ since $s_0  \notin \mathcal K'$ does not propagate any new weak promise. This leads to a contradiction. 

\item no $\neg x_i ^0$ is  weakened at any $\widetilde{l_{(j,k)}}$ in any $f_j$?  Let $s_0 = \widetilde{(l_{(j,k) }, x_i ^0)}$. Otherwise, let $\mathcal M$ a linear structure  of $\mathcal K' \cup\{ s_0 \} $ starting with the state $\widetilde {b_0}$, which never visit $\widetilde {x_i ^0}$, reaching the `weakened' state $s_0$ where $x_i ^0 \in l(s_0)$, crossing $\widetilde e_j$ and going directly to $\widetilde{(b_{i-1},\rho_i,\omega_i)}$. First,  $\neg b_i \vee \mathcal F (\beta))\, \mathcal U x_i ^0$ is fulfilled at $s_0$. Second, from $\widetilde {(\beta, \delta_3,\gamma_3)} $ , no constraint enable to propagate $\mathcal F (\beta)$, and any others remaining promises of $f-promise$ are  fulfilled at  $(\widetilde{f-promises \setminus \{\beta \},\delta_2,\gamma_2})$. From $(\widetilde{f-promises \setminus \{\beta \},\delta_2,\gamma_2})$ any constraint is obviously checked. Furthermore,  $s_0  \notin \mathcal K'$, thus it does not propagate any new weak promise. Thus, $\mathcal M$ is a model of $MU_0(Temp(\Phi))$.   This leads to a contradiction. 

 \item no $\neg a_{i -1}$ is  weakened at any $\widetilde{l_{(j,k)}}$ in any $f_j$? Let $s_0 = \widetilde{(l_{(j,k)} , a_{i-1})}$. Otherwise, Let $\mathcal M$ a linear structure of $\mathcal K'  \cup\{ s_0\} $ starting with the state $\widetilde {b_0}$,  reaching $\widetilde{x^0 _{i}}$, displays the `weakened' state $s_0$ where $a_{i -1} \in l(s_0)$, crosses $\widetilde e_j$   and goes straightforward to the solely branch where  $\widetilde{(x_{i} ^0 ,\mu_i,\nu_i)}$ occurs. First, the propagated $(\neg x^0_i \vee \mathcal F (\beta))\mathcal Wa_{i-1}$ is fulfilled at $s_0$. Second, from $\widetilde {(\beta, \delta_3,\gamma_3)} $ , no constraint enable to propagate $\mathcal F (\beta)$, and any others remaining promises of $f-promise$ are  fulfilled at  $(\widetilde{f-promises \setminus \{\beta \},\delta_2,\gamma_2})$. From $(\widetilde{f-promises \setminus \{\beta \},\delta_2,\gamma_2})$ any constraint is obviously checked. Furthermore,  $s_0  \notin \mathcal K'$, thus it does not propagate any new weak promise.   Thus, $\mathcal M$ is a model of $MU_0(Temp(\Phi))$.  This leads to a contradiction.

 \item no $\neg b_{i}$ is  weakened at any $\widetilde{l_{(j,k)}}$ in any $f_j$? Let $s_0 = \widetilde{(l_{(j,k)} , b_i)}$. Otherwise, Let $\mathcal M$ a linear structure of $\mathcal K' \cup\{s_0\} $ starting with the state $\widetilde {b_0}$, which never visit $\widetilde {x_i ^0}$ and that reach $\widetilde{(b_{i-1},\rho_i,\omega_i)}$. First, the propagated $(g \vee \mathcal F (\beta)) \mathcal W b_{i}$ is fulfilled at $s_0$. Second, from $\widetilde {(\beta, \delta_3,\gamma_3)} $ , no constraint enable to propagate $\mathcal F (\beta)$, and any others remaining propagated promises of $f-promise$ are  fulfilled at  $(\widetilde{f-promises \setminus \{\beta \},\delta_2,\gamma_2})$. From $(\widetilde{f-promises \setminus \{\beta \},\delta_2,\gamma_2})$ any constraint is obviously checked. Furthermore,  $s_0  \notin \mathcal K'$, thus it does not propagate any new weak promise.  Thus, $\mathcal M$ is a model of $MU_0(Temp(\Phi))$.  This leads to a contradiction.

\item Let $\mathcal M$ a $\mathcal K '$  linear structure  and  $j_0 \notin H$. But, if $\mathcal M  \vDash (MU_0(Temp(\Phi))) [Occ(l_{(j_0,k)})\leftarrow TRUE]$, then let $\mathcal  M_{H,j_0}=  \mathcal M[\widetilde{l_{(j,k')} }\leftarrow erase ][\widetilde{e_j }\leftarrow erase ][\widetilde{d_j }\leftarrow erase ]_{ \{(j,k') | j \in  H \cup \{j_0 \} , l(j,k') \in C_j\}}$. According that neither $\neg a_{i -1}$,$\neg b_{i}$ and $\neg x_i ^0$ are weakened at any state, $\mathcal  M_{H,j_0}$ propagates any of the triggered weak promises of types $.. \mathcal W b_{i}$, $...U x_i ^0$ or $... \mathcal W a_{i-1}$ in $ MU_0(Temp(\Phi))) [Occ(l_{(j_0,k)})\leftarrow TRUE]$ exactly as in $\neg \Psi'$, thus\footnote{As for point A of the (proof)} the states of $\mathcal  M_{H,j_0}$ actually lies (if one renames the variables) in $\mathcal K_{\Phi[C_j\leftarrow TRUE]_{j \in  H\cup \{j_0 \} }} \subset \mathcal K_{\Phi[C_j\leftarrow TRUE]_{j \in  H \cup \{j_0 \}}}'$. Furthermore, remark that since $\neg \beta$ is not weakened too, then any fulfilling state from $\mathcal M$ is still in $\mathcal M_{H,j_0}$. By renaming the remaining index $j'$ in $\mathcal  M_{H,j_0}$ and projecting over the propositional variables of $Temp(\Phi[C_j\leftarrow TRUE]_{j \in  H \cup \{j_0 \}}))$, we have $\mathcal  M_{H,j_0}\vDash (Temp(\Phi[C_j\leftarrow TRUE]_{j \in  H \cup \{j_0 \}}))$  because  the weak promises of $Temp(\Phi[C_j\leftarrow TRUE]_{j \in  H \cup \{j_0 \}})$ are still   eventually fulfilled\footnote{If the promise is fulfilled at $MU_0(Temp(\Phi))$} at the remaining states  $\widetilde {b_i}_{|j_0}$, $\widetilde {a_i}_{|j_0}$ or $\widetilde {x_i^0}_{|j_0}$ ($\mathcal F (\beta)$ cannot be triggered because it would not be fulfilled in $\mathcal K'_{|j_0}$). Thus, $(\Phi[C_j\leftarrow TRUE]_{j \in  H \cup \{j_0 \}})$ is satisfiable for any $j_0$, since $MU_0(Temp(\Phi))$ is a MU.

\item Assume  $(\Phi[C_h\leftarrow TRUE]_{j \in  H })$ is satisfiable, then $Temp(\Phi[C_j\leftarrow TRUE]_{j \in  H }) $ is satisfiable with a model $\mathcal  M_{H}$  ranging over $\mathcal K_{\Phi[C_j\leftarrow TRUE]_{j \in  H }}$. By renaming the index $j$ let $\mathcal  M =   \mathcal  M_{H}$ $(\widetilde b_0),\mathcal  M_{H}$ $(\widetilde b_1),..., $ $\mathcal  M_{H}(\widetilde{d_j }),   \mathcal  M_{H}(\widetilde{\emptyset}) , \mathcal  M_{H}(\widetilde{e_j })...$ which is $\mathcal  M_{H}$ into which some states have been added, some propositional variables have been added, such that the resulting linear structure is almost in $\mathcal K$. Precisely,  if any    $\widetilde \emptyset $ is substituted in $\mathcal  M $ by the corresponding $\widetilde l_{(j,k)}$( weakened at $MU_0(Temp( \Phi)$, so $ j \in H$ ) the modified linear structure is now a $\mathcal K$-linear structure . Since, $\widetilde \emptyset \notin \mathcal K'$, $\mathcal  M $ does not trigger any new weak promise from $MU_0(Temp(\Phi))$ w.r.t. $Temp(\Phi[C_h\leftarrow TRUE]_{j \in  H }) $. Furthermore, since any propagated weak promise which is fulfilled   of $Temp(\Phi[C_h\leftarrow TRUE]_{j \in  H }) $,  is also fulfilled for $MU_0(Temp(\Phi))$, and since the remaining consistency part of $MU_0(Temp(\Phi))$ never triggers a $\mathcal F (\beta)$ because there is no visit at a state where $l_{(j,k)}$ holds for $j \in H$, then  $\mathcal M \vDash MU_0(Temp(\Phi))$. But this is a contradiction since $MU_0(Temp(\Phi))$ is unsatisfiable, i.e., $(\Phi[C_h\leftarrow TRUE]_{h \in H})$ can solely be FALSE.
\end{itemize}

\end{enumerate}

Since the FPSPACE complete MF search problem is Logspace reducible to the MU for LTL search problem, the FPSPACE hardness is shown.

 Consider now the Inherent Non Vacuity (INVac) search problem, i.e. searching a limit strengthening $h$ of $g$  such that $h \equiv g$. Let $g=\neg [(Temp(\Phi)\wedge u) \vee (\neg Temp(\Phi) \wedge \neg u)]$ . An INVac strenghtening $h$ of $g$ is such that $u$ is at TRUE in $h$ iff $Temp(\Phi)$ is unsatisfiable. Furthermore in this case,  the weakening of $Temp(\Phi)$ at $h$ is minimal, i.e. it is in $MU(Temp(\Phi))$. We deduce, the following corollary.
\begin{corollary}($INVac_{Occ(sf)}$-search) Given a LTL formula $f$, searching an inherent non vacuous strengthening of $f$  (for single occurrences) is FPSPACE-complete\end{corollary}

\section{Conclusion}



We  have shown that  the  MF   search problem, the (LTL) MU search problem  and the (LTL) INVac search problem are  FPSPACE Complete. Furthermore, we have shown that the MU-dec and the Inherent Vacuity checking decision problem are PSPACE complete.     Although deficiency is the Backbone of lower Complexity Bound in the QBF case \cite{BuningZ08},  no corresponding bound exists for LTL. 


\bibliographystyle{plain}

\bibliography{bibliofinal}

\begin{thebibliography}{10}

\bibitem{SistlaC85}
A.P.Sistla and E.Clarke.
\newblock The complexity of prop. lin. temp. logics.
\newblock {\em J. ACM1985}.

\bibitem{SistlaVW87}
A.P.Sistla, M.~Vardi, and P.~Wolper.
\newblock The complementation problem for b{\"u}chi automata with appplications
  to temporal logic.
\newblock {\em Theor. Comput. Sci. 1987}.

\bibitem{ArmoniFFGPTV03}
Armoni and all.
\newblock Enhanced vacuity detection in ltl.
\newblock In {\em CAV}, 2003.

\bibitem{BeerBER01}
I.~Beer, S.~Ben-David, C.~Eisner, and Y.~Rodeh.
\newblock Efficient detection of vacuity in temporal model checking.
\newblock {\em FMSD 2001}.

\bibitem{BeerBCOT09}
Ilan Beer, Shoham Ben-David, Hana Chockler, Avigail Orni, and Richard~J.
  Trefler.
\newblock Explaining counterexamples using causality.
\newblock In {\em CAV}, pages 94--108, 2009.

\bibitem{BelovS11}
A.~Belov and JP.~M. Silva.
\newblock Minimally unsatisfiable boolean circuits.
\newblock In {\em SAT}, 2011.

\bibitem{BuningZ07a}
H.~B{\"u}ning and X.~Zhao.
\newblock An extension of deficiency and minimal unsatisfiability of quantified
  boolean formulas.
\newblock {\em JSAT}, 2007.

\bibitem{ChocklerHK08}
H.~Chockler, J.~Halpern, and O.~Kupferman.
\newblock What causes a system to satisfy a specification{\&}quest;.
\newblock {\em ACM Trans. Comput. Log.}, 9(3), 2008.

\bibitem{ChocklerS09}
Hana Chockler and Ofer Strichman.
\newblock Before and after vacuity.
\newblock {\em FMSD 2009}.

\bibitem{CimattiGS07}
A.~Cimatti, A.~Griggio, and R.~Sebastiani.
\newblock A simple and flexible way of computing small unsatisfiable cores in
  sat modulo theories.
\newblock In {\em SAT}, 2007.

\bibitem{CimattiRST07}
A.~Cimatti, M.~Roveri, V.~Schuppan, and S.~Tonetta.
\newblock Boolean abstraction for temporal logic satisfiability.
\newblock In {\em CAV}, pages 532--546, 2007.

\bibitem{Cook71}
Stephen~A. Cook.
\newblock The complexity of theorem-proving procedures.
\newblock In {\em STOC}, pages 151--158, 1971.

\bibitem{Papadimitriou1988}
C.Papadimitriou and W.David.
\newblock The complexity of facets resolved.
\newblock {\em JCSS.1988}.

\bibitem{DemriS98}
St{\'e}phane Demri and Ph. Schnoebelen.
\newblock The complexity of propositional linear temporal logics in simple
  cases (extended abstract).
\newblock In {\em STACS}, pages 61--72, 1998.

\bibitem{FismanKSV08}
D.Fisman and all.
\newblock A framework for inherent vacuity.
\newblock In {\em HVC2008}.

\bibitem{Emerson}
E.~Emerson.
\newblock Temporal and modal logic.
\newblock {\em HTCS, Volume B}, 1990.

\bibitem{Fainekos11}
Georgios~E. Fainekos.
\newblock Revising temporal logic specifications for motion planning.
\newblock In {\em ICRA}, pages 40--45, 2011.

\bibitem{Fisher91}
Michael Fisher.
\newblock A resolution method for temporal logic.
\newblock In {\em IJCAI}, 1991.

\bibitem{Hantry11F}
F.~Hantry and MS. Hacid.
\newblock Handling conflicts in depth-first search for ltl tableau to debug
  compliance based languages.
\newblock In {\em FLACOS}, 2011.

\bibitem{FleischnerKS02}
H.Fleischner, O.Kullmann, and S.Szeider.
\newblock Polynomial-time recognition of minimal unsatisfiable formulas with
  fixed clause-variable difference.
\newblock {\em Theor. Comput. Sci.}

\bibitem{BuningZ08}
H.K.B{\"u}ning and X.Zhao.
\newblock Computational complexity of quantified boolean formulas with fixed
  maximal deficiency.
\newblock {\em Theor. Comput. Sci.2008}.

\bibitem{BuningZ06}
H.K.B{\"u}ning and X.Zhao.
\newblock Minimal false quantified boolean formulas.
\newblock In {\em SAT}, 2006.

\bibitem{LynceM04}
I.Lynce and J.M.Silva.
\newblock On computing minimum unsatisfiable cores.
\newblock In {\em SAT2004}.

\bibitem{HalpernP01}
J.Y.Halpern and J.Pearl.
\newblock Causes and explanations: Part 1: Causes.
\newblock In {\em UAI}, 2001.

\bibitem{KupfermanV03}
O.~Kupferman and M.~Vardi.
\newblock Vacuity detection in temporal model checking.
\newblock {\em STTT}.

\bibitem{LiffitonS09}
M.~Liffiton and K.~Sakallah.
\newblock Generalizing core-guided max-sat.
\newblock In {\em SAT}, 2009.

\bibitem{Nadel10}
A.~Nadel.
\newblock Boosting minimal unsatisfiable core extraction.
\newblock In {\em FMCAD}, 2010.

\bibitem{Namjoshi04}
Kedar~S. Namjoshi.
\newblock An efficiently checkable, proof-based formulation of vacuity in model
  checking.
\newblock In {\em CAV}, pages 57--69, 2004.

\bibitem{PanV06}
G.~Pan and M.~Vardi.
\newblock Fixed-parameter hierarchies inside pspace.
\newblock In {\em LICS}, 2006.

\bibitem{papadimitriou}
C.~H. Papadimitriou.
\newblock {\em Computational Complexity}.
\newblock 1994.

\bibitem{Raman2011}
Vasumathi Raman and Hadas Kress-Gazit.
\newblock Analyzing unsynthesizable specifications for high-level robot
  behavior using ltlmop.
\newblock In {\em Proceedings of the 23rd international conference on Computer
  aided verification}, CAV'11, pages 663--668, Berlin, Heidelberg, 2011.
  Springer-Verlag.

\bibitem{Schuppan10}
V.~Schuppan.
\newblock Towards a notion of unsatisfiable cores for ltl.
\newblock {\em Sci.Comp.Pro.2010}.

\bibitem{SilvaL11}
JP.~M. Silva and I.~Lynce.
\newblock On improving mus extraction algorithms.
\newblock In {\em SAT}, 2011.

\bibitem{SimmondsDGC07}
J.~Simmonds, J.~Davies, A.~Gurfinkel, and M.~Chechik.
\newblock Exploiting resolution proofs to speed up ltl vacuity detection for
  bmc.
\newblock In {\em FMCAD}, pages 3--12, 2007.

\bibitem{StockmeyerM73}
Larry~J. Stockmeyer and Albert~R. Meyer.
\newblock Word problems requiring exponential time: Preliminary report.
\newblock In {\em STOC}, pages 1--9, 1973.

\bibitem{YuM05}
Yinlei Yu and Sharad Malik.
\newblock Validating the result of a quantified boolean formula (qbf) solver:
  theory and practice.
\newblock In {\em ASP-DAC}, pages 1047--1051, 2005.

\bibitem{ZM03}
L.~Zhang and S.~Malik.
\newblock Extracting small unsatisfiable cores from unsatisfiable boolean
  formula.
\newblock In {\em In Prelim of SAT’03}, 2003.

\end{thebibliography}



\end{document}